\documentclass[preprint,times,12pt]{elsarticle}

\pdfoutput=1

\usepackage{amssymb}
\usepackage[nodots]{numcompress}
\usepackage{booktabs}
\usepackage{graphicx}
\usepackage{caption}
\usepackage{subcaption}
\usepackage[section]{placeins}
\usepackage{dcolumn}
\usepackage{multirow}
\usepackage{hyperref}
\biboptions{sort&compress}
\newcolumntype{.}{D{.}{.}{1.1}}

\journal{}

\begin{document}

\begin{frontmatter}

\title{Characterizing the local primary dendrite arm spacing in directionally-solidified dendritic microstructures}

\author[arl,cavs]{M. A. Tschopp \corref{cor1}}
\cortext[cor1]{Corresponding Author}
\ead{mark.tschopp@gatech.edu}
\author[afrl]{J. D. Miller}
\author[cavs]{A. L. Oppedal}
\author[asu]{K. N. Solanki}

\address[arl]{Army Research Laboratory, Materials and Manufacturing Science Division, Lightweight and Specialty Metals Branch, APG, MD 21014}
\address[cavs]{Center for Advanced Vehicular Systems, Mississippi State University, Mississippi State, MS 39762}
\address[afrl]{Air Force Research Laboratory, Wright-Patterson AFB, OH 45433}
\address[asu]{School for Engineering of Matter, Transport, and Energy, Arizona State University, Tempe, AZ 85287}

\begin{abstract}

\small{
Characterizing the spacing of primary dendrite arms in directionally-solidified microstructures is an important step for developing process-structure-property relationships by enabling the quantification of (i) the influence of processing on microstructure and (ii) the influence of microstructure on properties. In this work, we utilized a new Voronoi-based approach for spatial point pattern analysis that was applied to an experimental dendritic microstructure. This technique utilizes a Voronoi tessellation of space surrounding the dendrite cores to determine nearest neighbors and the local primary dendrite arm spacing. In addition, we compared this technique to a recent distance-based technique and a modification to this using Voronoi tesselations. Moreover, a convex hull-based technique was used to include edge effects for such techniques, which can be important for thin specimens. These methods were used to quantify the distribution of local primary dendrite arm spacings, their spatial distribution, and their correlation with interdendritic eutectic particles for an experimental directionally-solidified Ni-based superalloy micrograph. This can be an important step for correlating with both processing and properties in directionally-solidified dendritic microstructures.
}
\end{abstract}


\end{frontmatter}

\section{Introduction}
Developing an enhanced understanding of mechanical behavior in materials relies upon sufficiently characterizing microstructure details at the relevant length scales that contribute to this behavior. Moreover, to truly enhance the predictive capability of processing-structure-property models that aim to improve material performance requires a quantitative stereological description of the relevant microstructure features and, thereby, the material itself.  Predictive models that effectively capture the linkage between processing and properties (through microstructure) can be utilized within an integrated computational materials engineering (ICME) approach to design materials and accelerate their insertion into application.

The focus of the present work is on single crystal nickel-based superalloys, which are used in turbine blades within the high temperature section of the modern turbine engine \citep{Ree2006, Pol2006}. In single crystal nickel-based superalloys, there are a number of length scales of microstructure that contribute to mechanical behavior, ranging from the $\gamma^\prime$ precipitates to pores and eutectic particles to the dendrites themselves.  At the largest microstructure length scale in directionally-solidified single crystal microstructures, the features of interest are the dendrites; many features at lower length scales (e.g., eutectic particles, precipitates, etc.) or at similar scales (e.g., porosity, freckle defects, etc.) are strongly associated with the dendrite arm spacing and morphology \citep{Whi2001,Ell2004,Mel2005,Lam2007,Bru2012}.    Historically, the primary dendrite arm spacing (PDAS) has been found to correlate with processing (e.g., solidification rate) \citep{McC1981,Hui2002,Wan2003, Mil2012,Bru2011,Bru2012} as well as with properties (e.g., creep strength, fatigue properties)\citep{Wil2008,Lam2007}.  For instance, Lamm and Singer \citep{Lam2007} produced single crystal nickel-based microstructures (PWA 1483) with a varied range of different dendrite arm spacings (250 $\mu$m to 600 $\mu$m) and found that decreasing the mean dendrite arm spacing was associated with an increased high-cycle fatigue life.  The fatigue cracks were found to originate at shrinkage porosity and the largest pores correlated with a large PDAS.

The traditional approach for measuring primary dendrite arm spacing in single crystal metals, whereby the number of dendrite cores in a specified area is related to the dendrite arm spacing \citep{Fle1974,Jac1976,McC1981} is given by:

\begin{equation}
\lambda = c \sqrt{\frac{A}{n}}
\label{lambda}
\end{equation}

\noindent where $\lambda$ is primary dendrite arm spacing, $A$ is the area analyzed, $n$ is the number of dendrites, and $c$ is a coefficient that depends on the microstructure.  McCartney and Hunt \cite{McC1981} showed that $c=0.5$ for a random array of points, $c=1$ for a square array of points, and $c=1.075$ for a hexagonal array of points; they had to apply a correction for the bulk dendrite arm spacing $\lambda$ as processing conditions caused a change in the local environment of the dendrites.  However, this approach is insufficient for capturing local arm spacings or the dendrite arm spacing distribution, and may provide problems with complex geometries such as turbine blades.  In fact, part of the motivation for quantifying the local PDAS is that a narrow distribution (i.e., low standard deviation) of local PDAS values may result in a more homogeneous distribution of interdendritic microstructure features and, more importantly, a narrow distribution of mechanical properties. 

The research objective herein is to evaluate the capability of some recent approaches, as well as some modified versions of these approaches, for characterizing the local dendrite arm spacing within experimental dendritic microstructures. In this work, an experimental dendritic microstructure is used for this analysis along with three different techniques that are based on the nearest neighbor spacing and/or a Voronoi tessellation of the dendrite cores. Comparison of existing and new metrics with traditional primary dendrite arm spacing metrics is discussed for both local and global measures. The current methods investigated supply statistical information of local spacing and coordination number while introducing a technique for addressing edge effects and examining the parameter sensitivity of these different methods.  In comparison to previous work \cite{Tsc2013}, this work introduces and compares the statistical distributions of local dendrite arm spacings for the four methods, for a more quantitative analysis.  It was found that augmenting existing techniques with Voronoi tesselations to define the subset of first nearest neighbors or refining existing Voronoi-based techniques resulted in a more physical description of the local dendrite arm spacing.  Moreover, for certain cases, the mean local dendrite statistics can adequately approximate the PDAS found with the traditional bulk characterization technique (Eq.~\ref{lambda}).

\section{Methodology}
\begin{figure}[hbt!]
        \centering
        \begin{subfigure}[b]{0.3\textwidth}
                \centering
                \includegraphics[width=\textwidth]{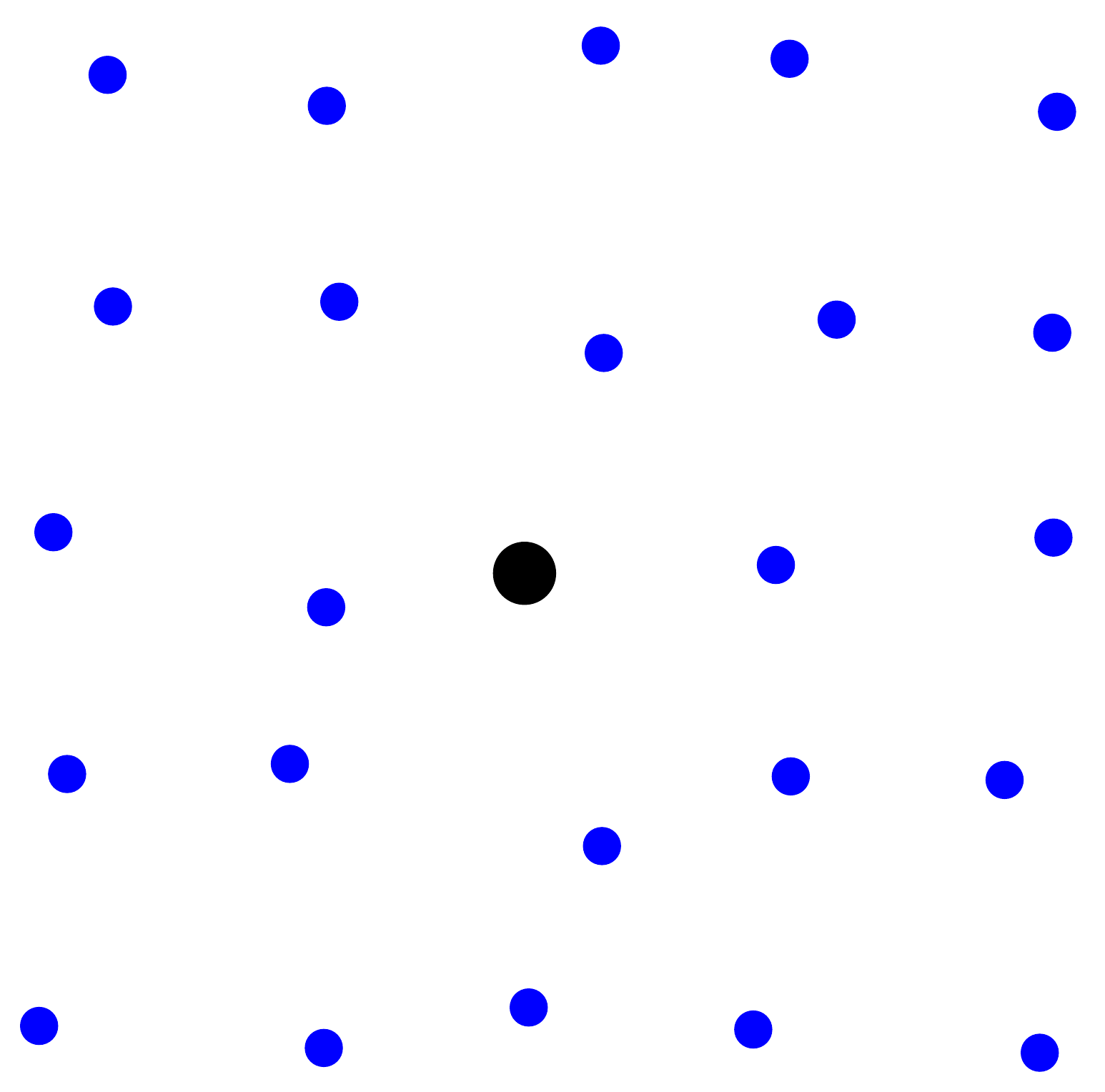}
                \caption{}
                \label{5x5}
        \end{subfigure}%
\quad
        \begin{subfigure}[b]{0.3\textwidth}
                \centering
                \includegraphics[width=\textwidth]{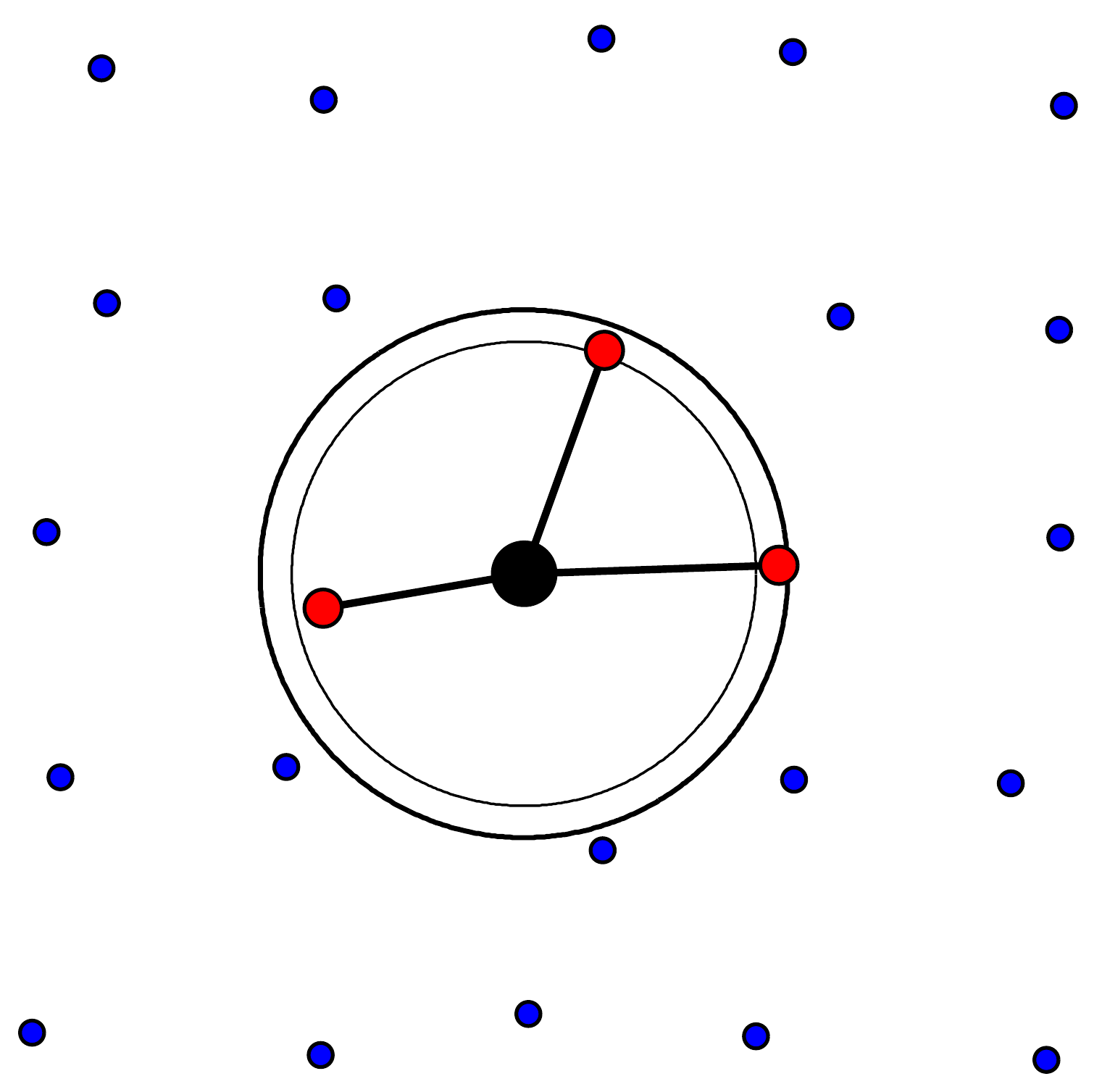}
                \caption{}
                \label{inner_circle}
        \end{subfigure}
\quad
        \begin{subfigure}[b]{0.3\textwidth}
                \centering
                \includegraphics[width=\textwidth]{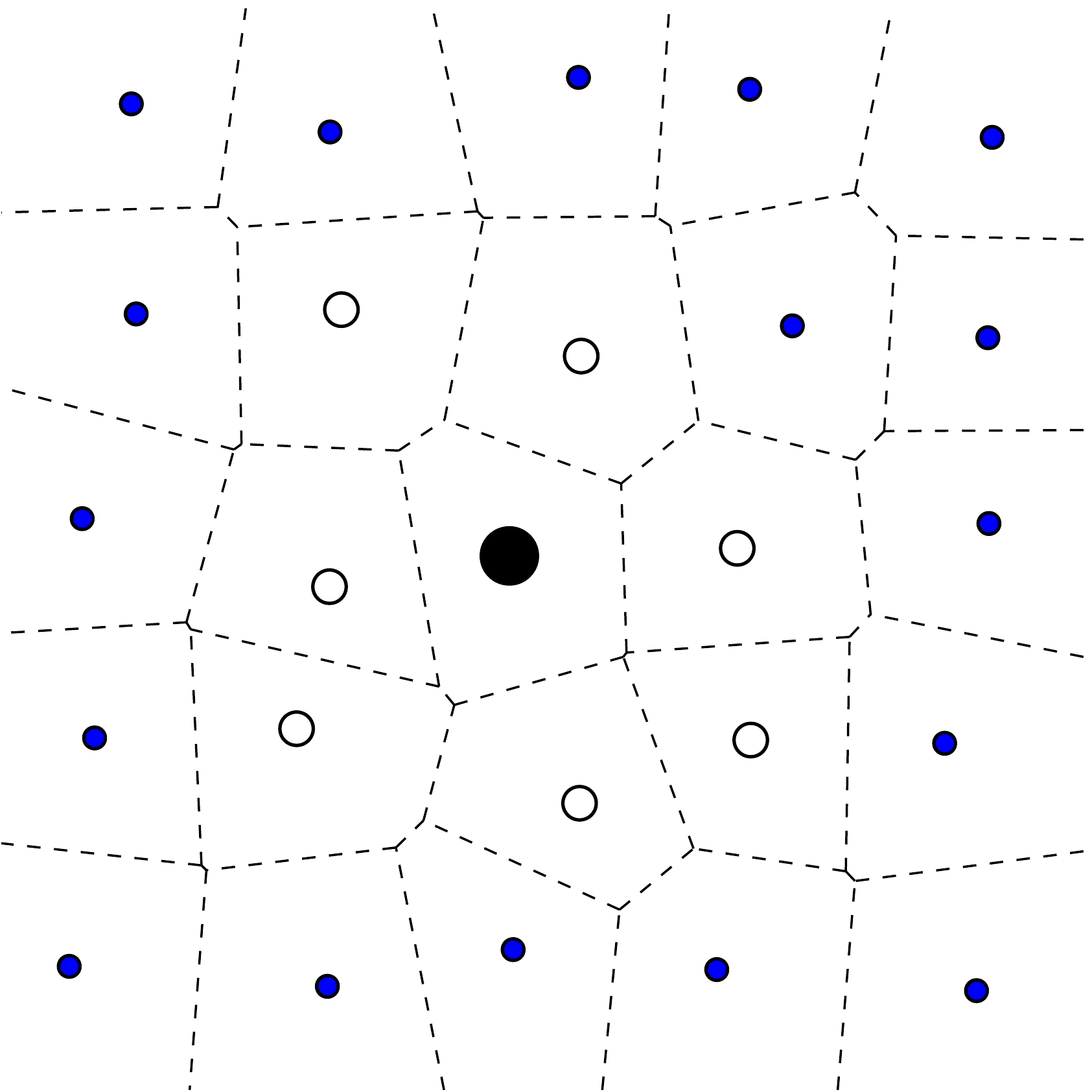}
                \caption{}
                \label{voronoi_tesselation}
        \end{subfigure}
\\
        \begin{subfigure}[b]{0.3\textwidth}
                \centering
                \includegraphics[width=\textwidth]{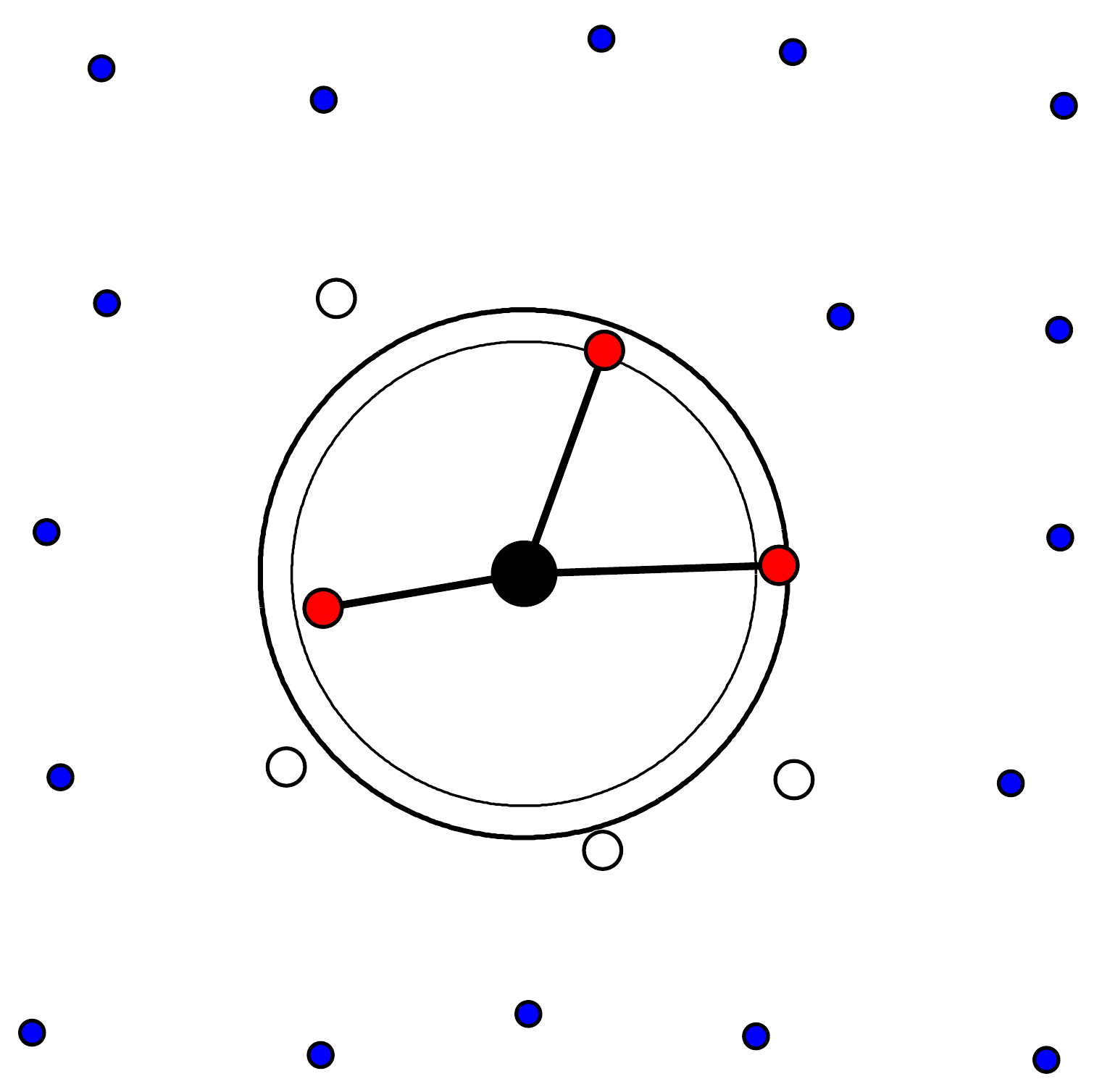}
                \caption{}
                \label{mod_warnken_reed}
        \end{subfigure}
\quad
        \begin{subfigure}[b]{0.3\textwidth}
                \centering
                \includegraphics[width=\textwidth]{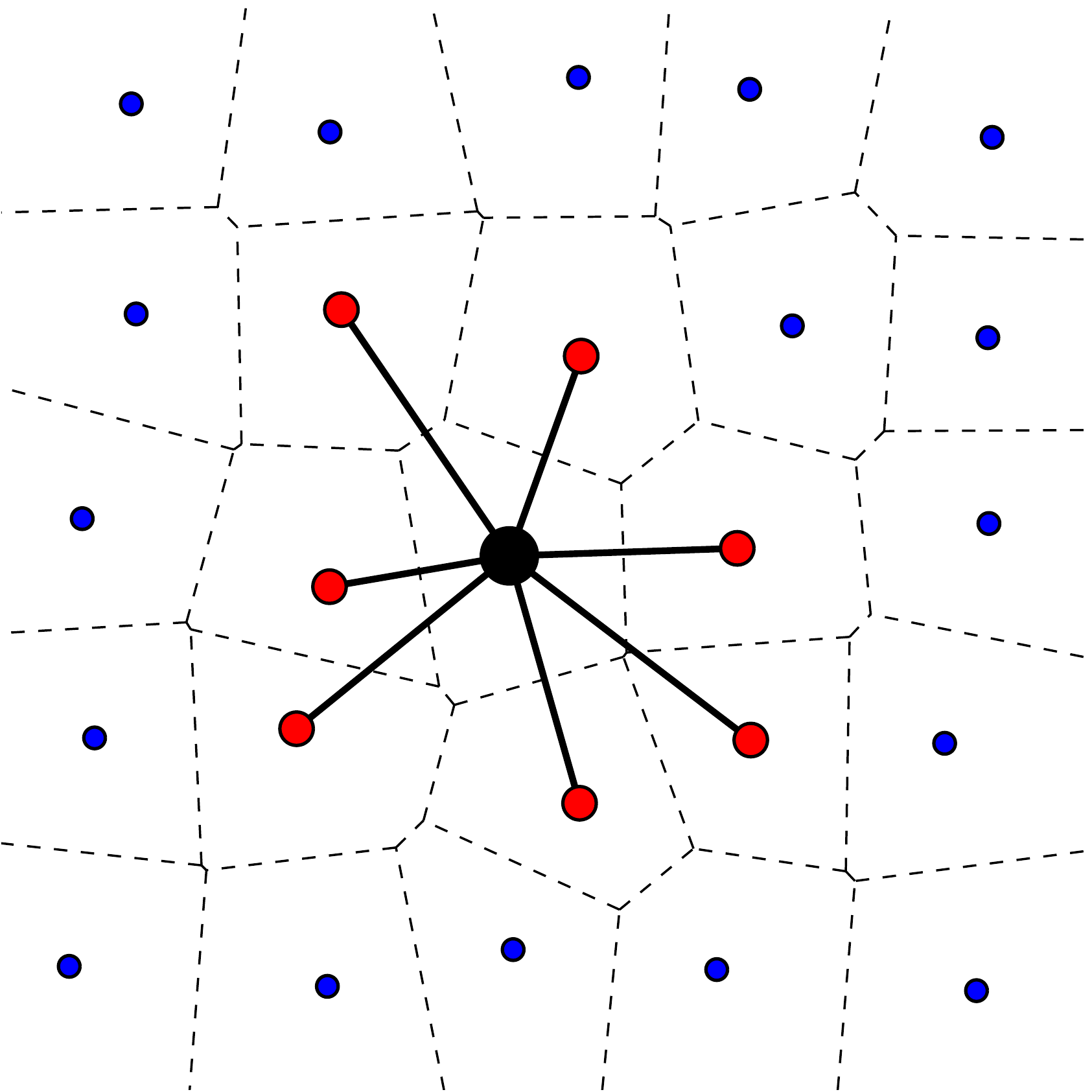}
                \caption{}
                \label{nearest_neighbor}
        \end{subfigure}
\quad
        \begin{subfigure}[b]{0.3\textwidth}
                \centering
                \includegraphics[width=\textwidth]{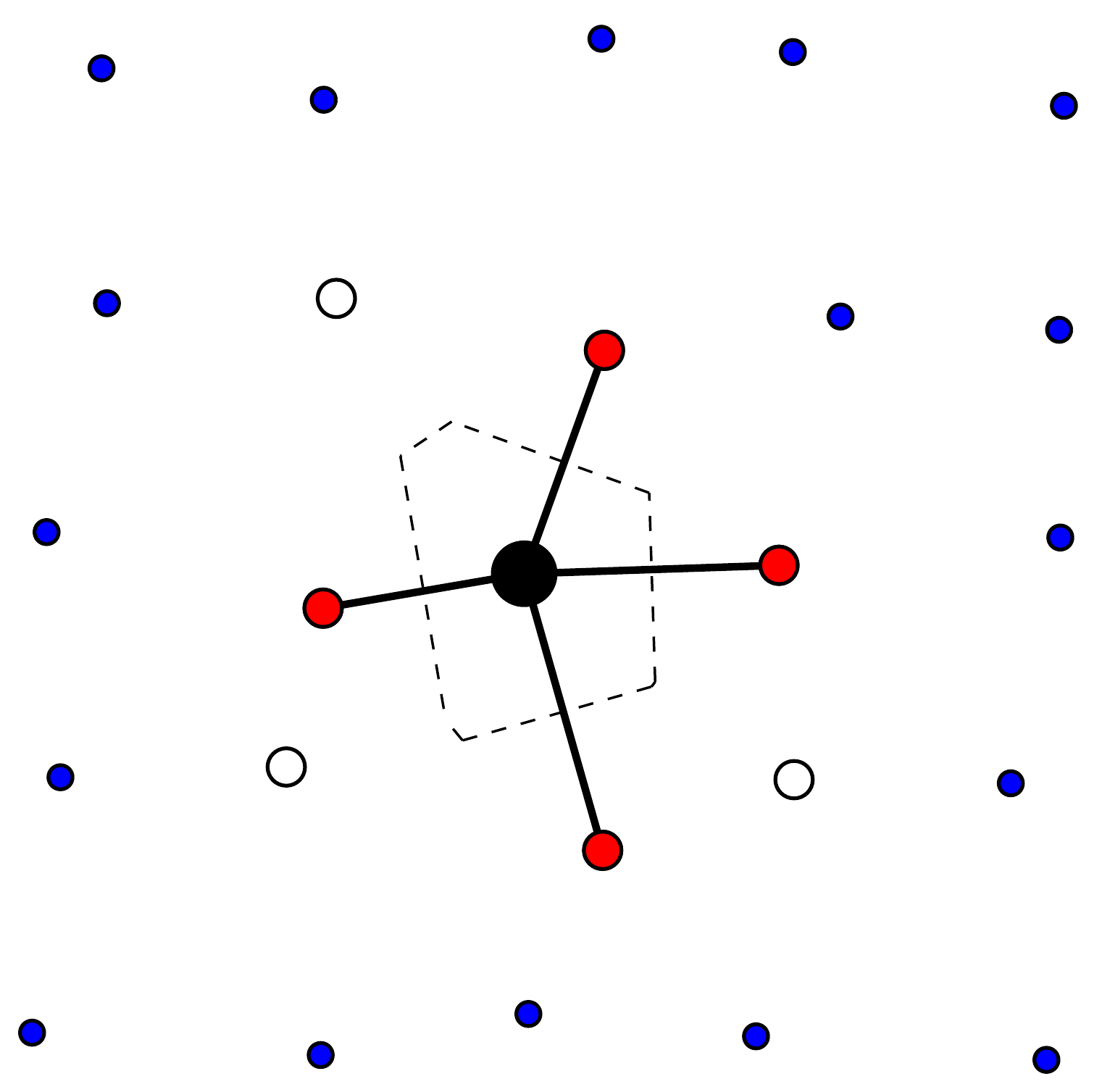}
                \caption{}
                \label{mod_tesselation}
        \end{subfigure}
        \caption{The difference between various methods for defining the nearest neighbors (red dots) and their spacing for a single point (large black dot). (a) Initial 5 x 5 cubic pattern with noise fraction of 100\% and noise level of 0.20$a_0$. (b) The Warnken--Reed method with $\alpha$ = 1.5 and $k_{initial} = 3$. The inner circle represents the average spacing, $d_{avg}$, of these neighbors and the outer circle represents the cutoff for adding the next neighbor, $d_{avg} + \alpha d_{std}$. (c) Voronoi tesselation diagram for the points. The potential first nearest neighbors are identified through shared vertices with each point. (d) The modified Warnken--Reed method with $\alpha = 1.5$ and $k_{initial} = 3$, whereby the neighbors are restricted to only those identified using the Voronoi tesselation. (e) Using only shared vertices (and connecting lines forming a polygon) of the Voronoi tesselation to identify the nearest neighbors  ($d_{crit}=0.0$). (f) Modified tesselation-based technique whereby the nearest neighbors are identified as those with line lengths above a critical threshold fraction of the total perimeter line length $d_{crit}=0.10$ of the tesselated polygon for the point (Reprinted from \cite{Tsc2013}).}
\label{various_methods}
\end{figure}

The approach utilized herein to measure the local dendrite arm spacing is based on a Voronoi tessellation of the spatial array of dendrite cores.  The following analysis techniques were implemented in MATLAB R2012a (The MathWorks, Inc.).  To illustrate how the present method works and differs from some other published methods, we generated a synthetic 5x5 cubic pattern of points with a small degree of noise (100\% noise fraction, 0.20$a_0$ noise fraction [2]), as shown in Figure~\ref{5x5} \cite{Tsc2013}. For the purposes of describing several different methods shown in Figure~\ref{various_methods}, this synthetic pattern of points can be considered as the cores of primary dendrites.

One such method for measuring the local dendrite arm spacing is the Warnken--Reed method \citep{War2011,War2011a}. The Warnken--Reed method calculates the dendrite arm spacing for a single point (black dot) by starting with an initial number of nearest neighbors (3 closest neighbors) and iteratively adding potential nearest neighbors that are within a cutoff distance defined by the already-added nearest neighbors. For instance, the inner circle in Figure~\ref{inner_circle} represents the average spacing, $d_{avg}$, of these neighbors and the outer circle represents the cutoff for adding the next neighbor, $d_{avg} + \alpha d_{std}$, where $d_{std}$ is the standard deviation of the nearest neighbor spacings and $\alpha$ is a parameter that is typically between 1 and 2. Neighbors continue to be added until the cutoff does not include any new neighbors. The local coordination number and dendrite arm spacing is calculated from the neighbors added (shown as red dots). However, if the standard deviation of the distances of the nearest neighbors $d_{std}$ or the parameter $\alpha$ is large, this technique can continue to add nearest neighbors beyond the first nearest neighbors; our implementation stopped after 20 nearest neighbors. Clearly, a method for restricting the number of nearest neighbors using such a technique is necessary.

A simple way of identifying the potential first nearest neighbors is to perform a Voronoi tessellation of the space surrounding the points, as shown in Figure~\ref{voronoi_tesselation}.  The polygon edges are equidistant between the points contained in the two adjacent polygons and the triple points (merging of three lines) are equidistant between the points contained in the three adjacent polygons.  Therefore, the first nearest neighbors (FNNs, shown as open circles in Fig.~\ref{voronoi_tesselation}) correspond to the edges of the central polygon (that contains the black dot).  This subset of points is the maximum number of nearest neighbors that the central point can have.  

In this manner, several techniques have been identified to quantify a local dendrite arm spacing based on the Voronoi-identified FNNs \cite{Tsc2013}. For instance, the Voronoi Warnken--Reed method (Figure~\ref{mod_warnken_reed}) only includes the Voronoi FNNs as potential nearest neighbors and cannot expand beyond these, alleviating a potential problem of selecting second nearest neighbors or greater. Another method using the Voronoi FNNs is to consider all of these potential nearest neighbors as nearest neighbors (Figure~\ref{nearest_neighbor}), as in Brundidge et al.~\citep{Bru2011}. Unfortunately, this approach is sensitive to small perturbations in the spatial positions of the neighbors. For instance, if the lower right hand neighbor in Figure~\ref{nearest_neighbor} moves away from the central point, it no longer shares an edge with the polygon containing the black dot; in this scenario, the two adjacent polygons on either side effectively ``pinch'' off this neighbor. This scenario, however, has a physical basis as these two dendrite cores mainly compete with the central core, and the lower right core has a much less prominent effect on the central core. The last method, which is examined in the present paper, utilizes a criterion based on the edge lengths of the Voronoi polygon. In Figure~\ref{mod_tesselation}, those neighbors with edge lengths less than a critical fraction, $d_{crit}$, of the total polygon perimeter are excluded as nearest neighbors (e.g., 10\% in Figure~\ref{mod_tesselation}). In the present study, the local dendrite arm spacing statistics are evaluated using these four techniques: Warnken--Reed, Voronoi Warnken--Reed, and the Voronoi technique with ($d_{crit}>0$) and without ($d_{crit}=0$) a line length threshold.

\begin{figure}[hbt!]
        \centering
        \begin{subfigure}[b]{0.3\textwidth}
                \centering
                \includegraphics[width=\textwidth]{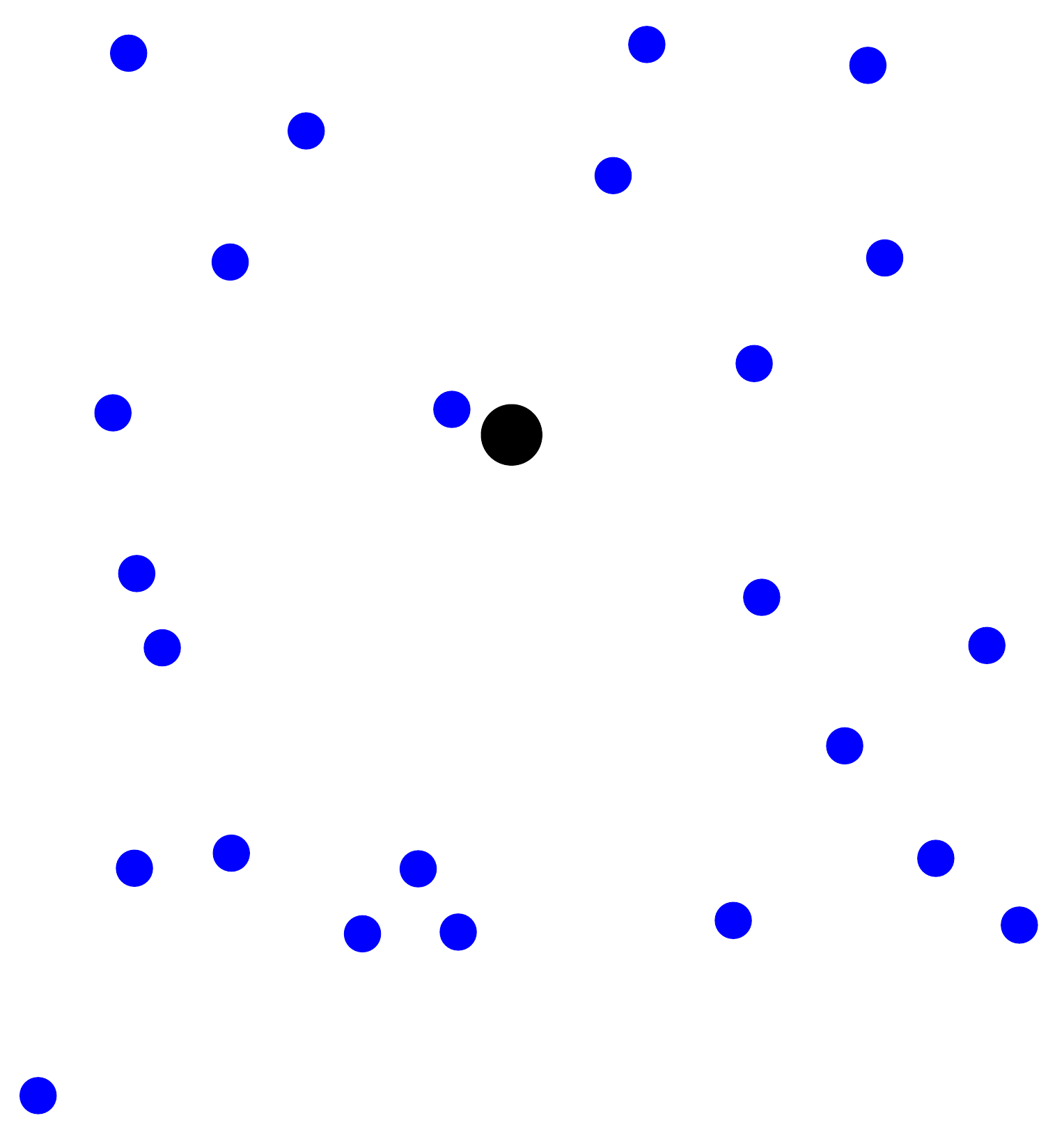}
                \caption{}
                \label{5x52}
        \end{subfigure}%
\quad
        \begin{subfigure}[b]{0.3\textwidth}
                \centering
                \includegraphics[width=\textwidth]{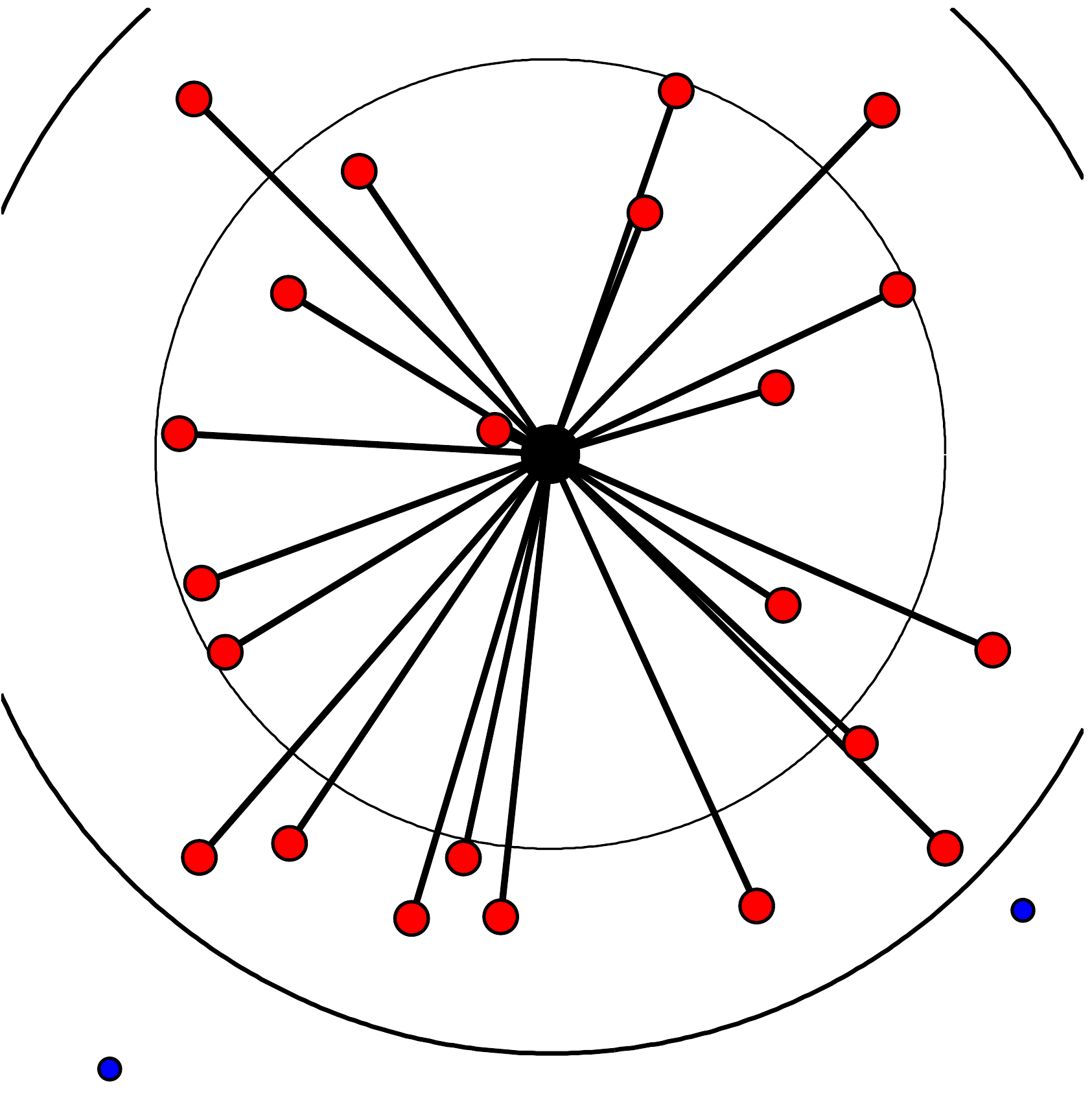}
                \caption{}
                \label{inner_circle2}
        \end{subfigure}
\quad
        \begin{subfigure}[b]{0.3\textwidth}
                \centering
                \includegraphics[width=\textwidth]{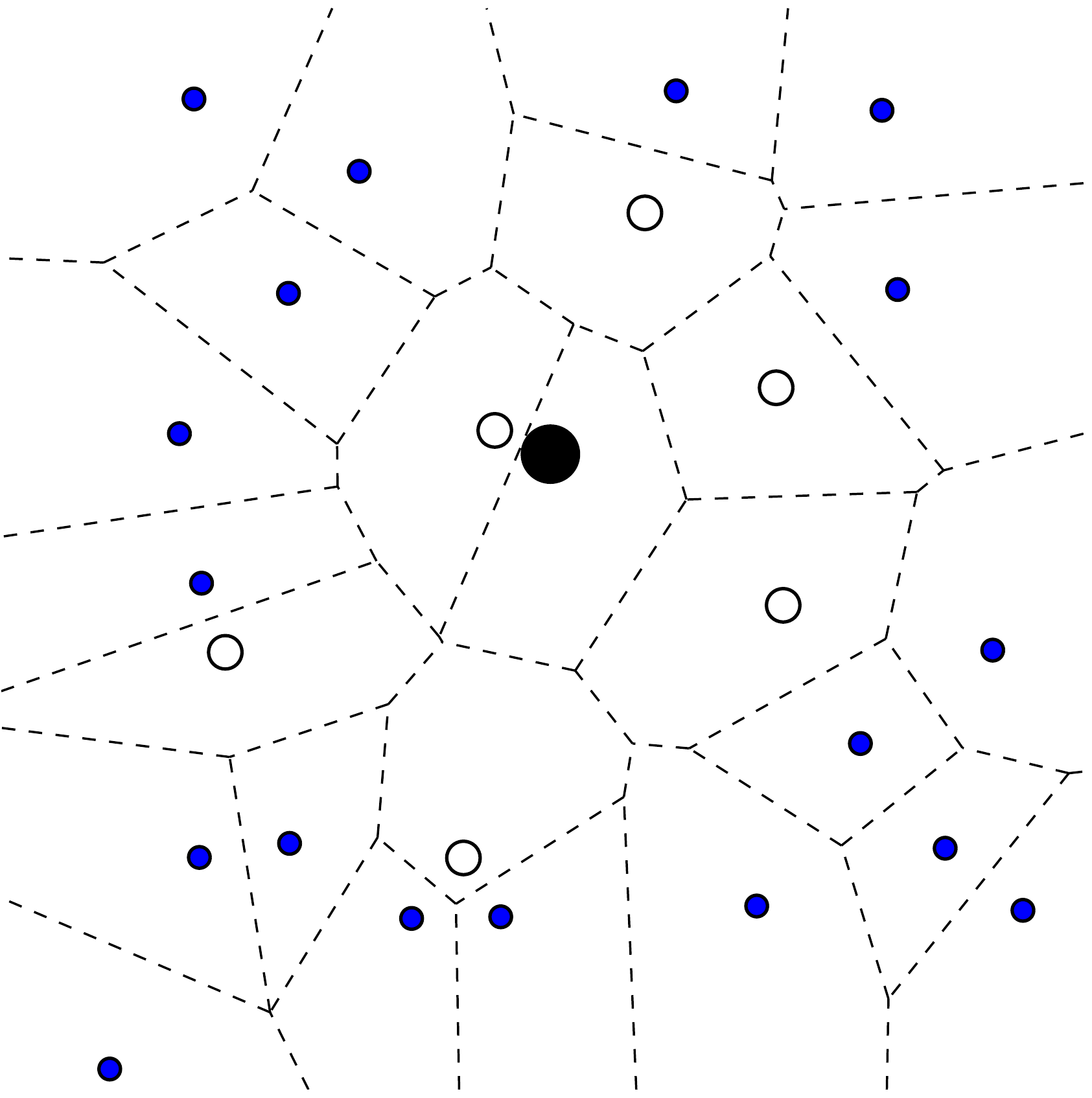}
                \caption{}
                \label{voronoi_tesselation2}
        \end{subfigure}
\\
        \begin{subfigure}[b]{0.3\textwidth}
                \centering
                \includegraphics[width=\textwidth]{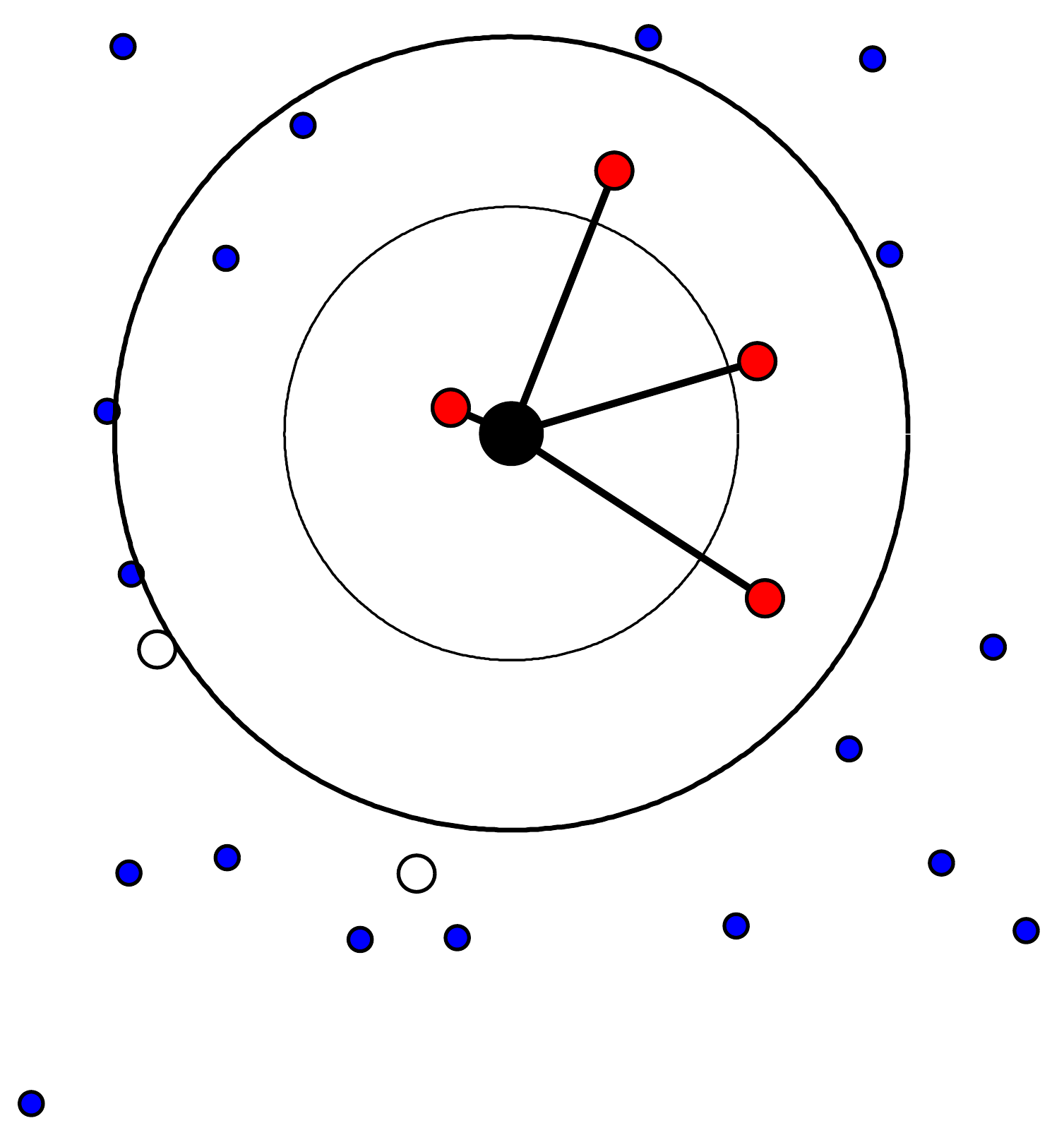}
                \caption{}
                \label{mod_warnken_reed2}
        \end{subfigure}
\quad
        \begin{subfigure}[b]{0.3\textwidth}
                \centering
                \includegraphics[width=\textwidth]{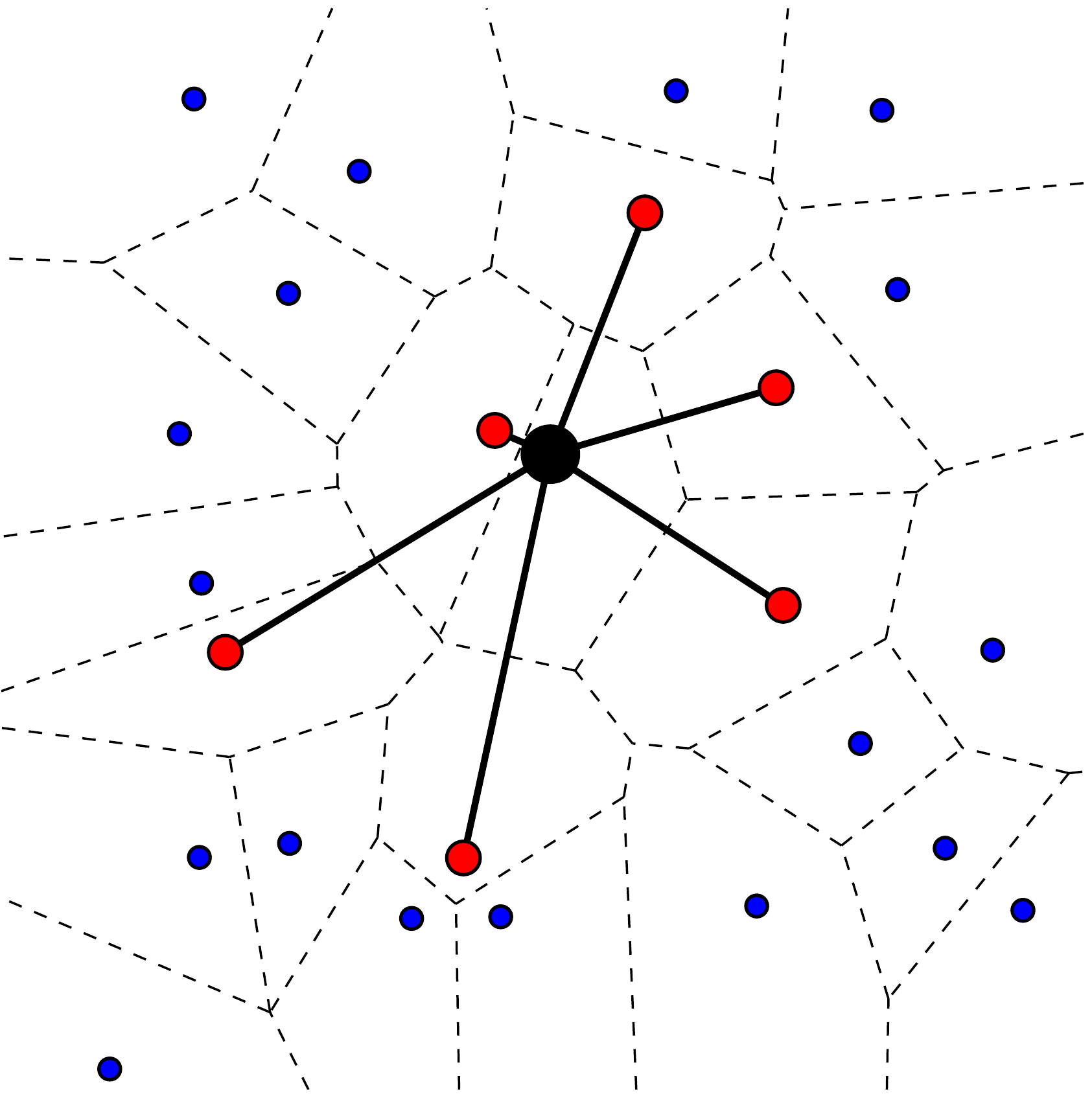}
                \caption{}
                \label{nearest_neighbor2}
        \end{subfigure}
\quad
        \begin{subfigure}[b]{0.3\textwidth}
                \centering
                \includegraphics[width=\textwidth]{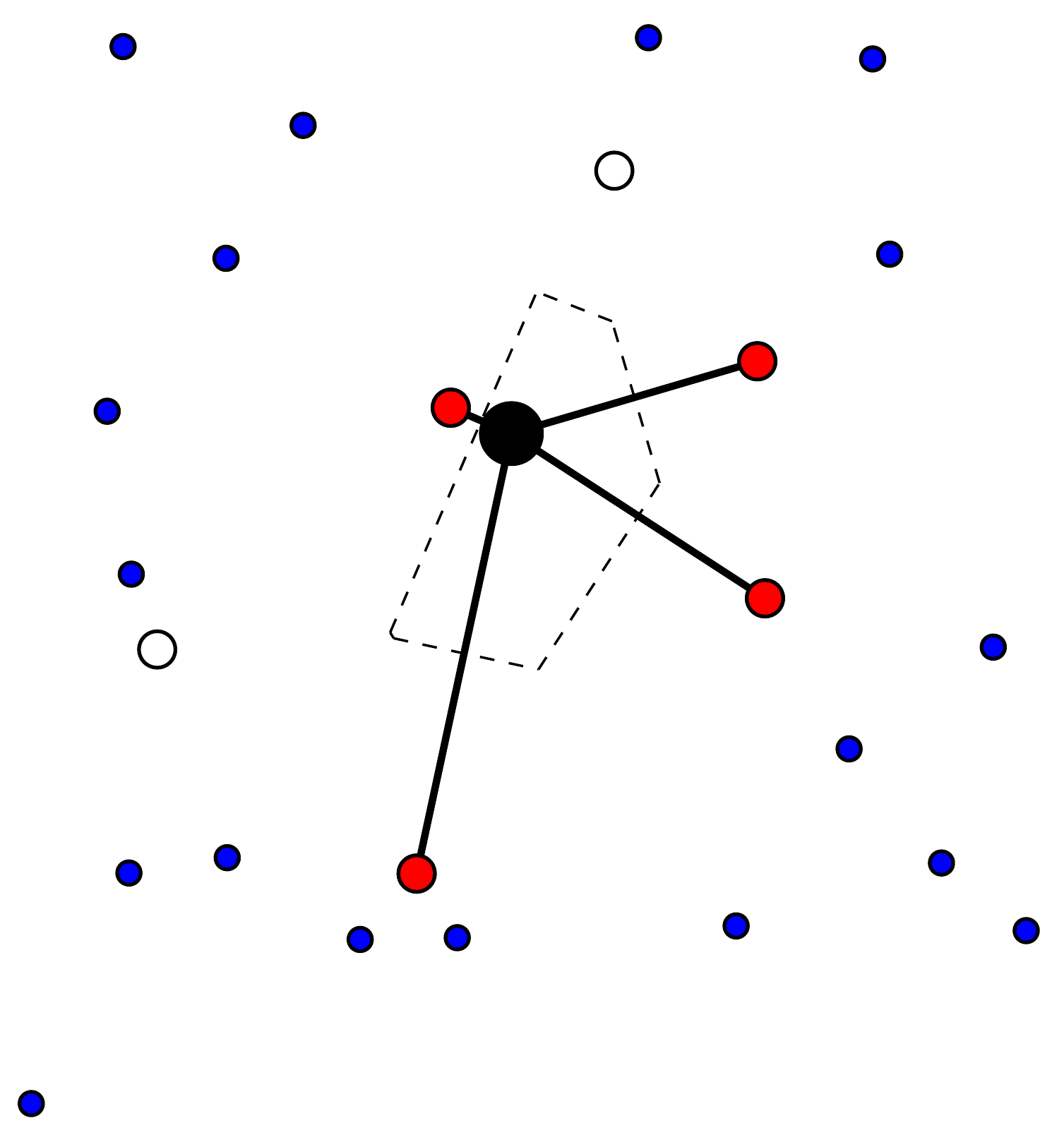}
                \caption{}
                \label{mod_tesselation2}
        \end{subfigure}
        \caption{The difference between various methods for defining the nearest neighbors (red dots) and their spacing for a single point (large black dot) with a distorted local environment.  Parts (a)-(f) are as Figure 1: (a) initial pattern, (b) the Warnken--Reed method with $\alpha$ = 1.5 and $k_{initial} = 3$, (c) the Voronoi tesselation diagram for the points, (d) the Voronoi-modified Warnken--Reed method with $\alpha = 1.5$ and $k_{initial} = 3$, (e) the Voronoi method ($d_{crit}=0.0$), and (f) the modified tesselation-based technique with $d_{crit}=0.10$.}
\label{various_methods2}
\end{figure}

As an example of a more disordered structure, Figure~\ref{various_methods2} plots the four different methods for a different configuration of surrounding points (dendrite cores).  In Figure~\ref{inner_circle2}, the iterative Warnken--Reed method continues to non-physically add neighbors beyond the first nearest neighbors due to a large initial $d_{std}$ value from the initial three distances.  The Voronoi-modified version in Figure~\ref{mod_warnken_reed2} stops at four nearest neighbors despite the fact that several points lie within the outer boundary computed by this method.  The Voronoi method with $d_{crit}=0.0$ clearly overestimates the number of nearest neighbors, while the four nearest neighbors identified through $d_{crit}=0.10$ (Figure~\ref{mod_tesselation2}) perhaps offers a better approximation of the number of nearest neighbors.  Interestingly, comparing the methods in Figure~\ref{mod_warnken_reed2} and \ref{mod_tesselation2}, the coordination number is the same, but the nearest neighbors identified is different.  This is due to the Warnken--Reed method being a distance-based method, and identifying the four closest neighbors, while the modified Voronoi technique is based on the edge lengths of the Voronoi polygon, and hence utilizes this to identify nearest neighbors (which may not be the closest neighbors). 

The traditional PDAS metric does not consider the order or disorder of the dendrites within the microstructure.  Figure~\ref{various_methods2} illustrates why a local metric for PDAS may be needed.  For the field of view given in Figs.~\ref{5x5} and \ref{5x52}, the bulk PDAS metric would be the same since the number of dendrites $n$ and the area $A$ are equal (see Eq.~\ref{lambda}).  However, the disorder of the dendritic structure in the case of Fig.~\ref{various_methods2} may yield (i) a more uneven distribution of solute elements, (ii) the formation of second phase particles, (iii) the formation of gas or shrinkage porosity, or (iv) the lateral growth of secondary dendrite arms.  Hence, in addition to the bulk PDAS values, understanding how processing conditions may impact the disorder of the dendritic structure may be important for understanding the properties of directionally-solidified alloys.

\begin{figure}[bht!]
\centering
\includegraphics[width = 0.7\textwidth]{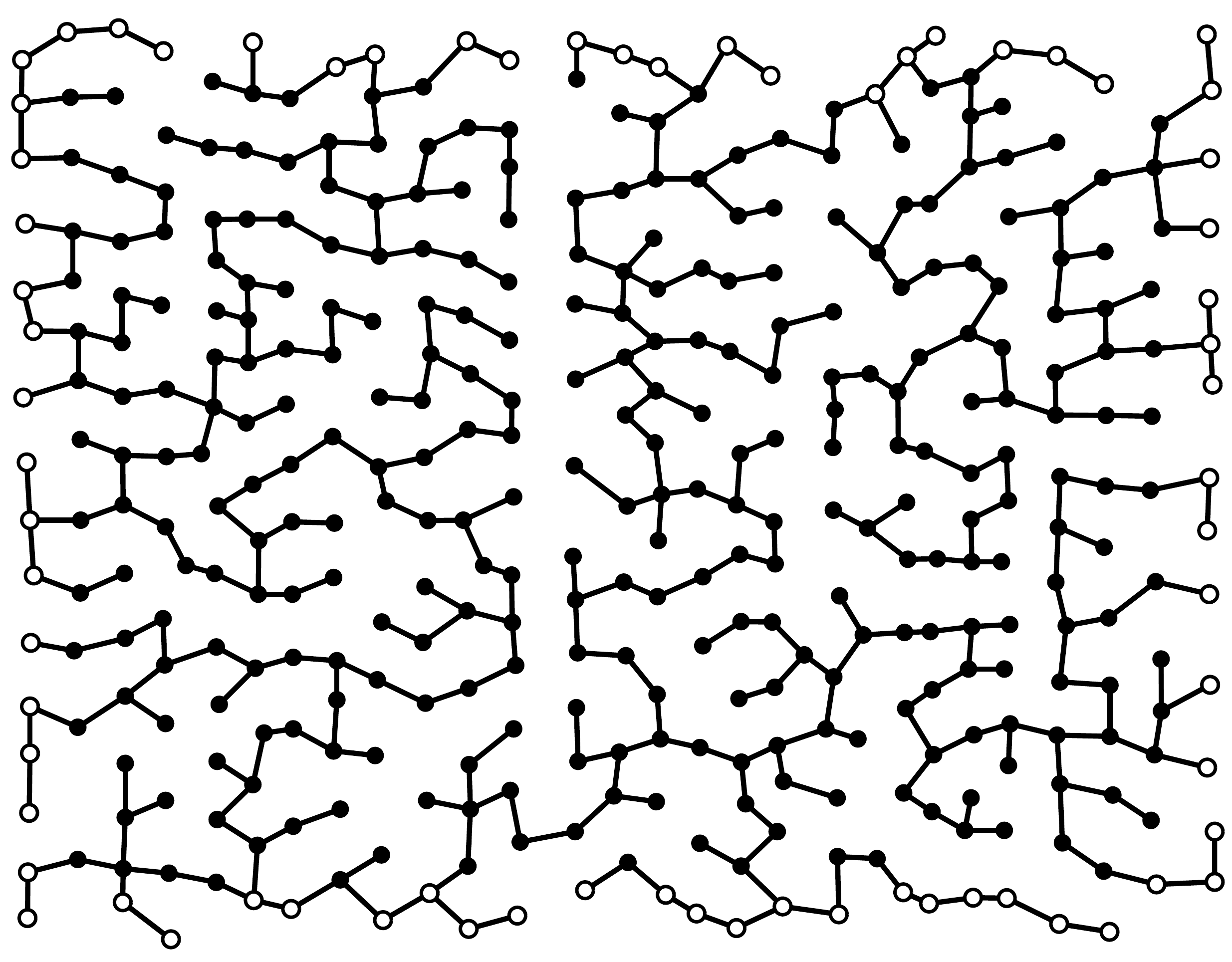}
\caption[]{Minimal spanning tree method \citep{Dus1986} for defining the spacing and homogeneity of a microstructure (set of points), whereby the lines represent the minimal distance of connecting line segments.  The set of points selected for this example were selected from the dendrite cores shown in Figure \ref{sx_nickel}, where the white dots indicate `edge' dendrite cores. }
\label{MST}
\end{figure}

Other techniques exist for quantifying the homogeneity or heterogeneity of primary dendrite arm spacing in directionally-solidified dendritic microstructures.  For instance, the minimal spanning tree (MST) method \citep{Dus1986} provides a statistical analysis of the disorder in a system of points by connecting all points with the shortest line segments possible.  In this manner, the mean distance of all line segments ($m$) and the standard deviation ($\sigma$) characterize the disorder of the system and casting these values into a $m$-$\sigma$ design space allows for comparison between different point systems \citep{Dus1986}.  This has been effectively applied to characterize the mean dendrite arm spacing, PDAS distribution, and the disorder in first Pb-Tl alloys \citep{Bil1991} and subsequently in other alloy systems \citep[e.g.,][]{Tew2002,Hui2002,Pen2013}.  As an example of this technique, Figure \ref{MST} plots the dendrite cores and connecting line segments for the single crystal nickel-based superalloy micrograph used in this study (Figure \ref{sx_nickel}).  Moreover, other methods such as radial distribution functions, fast Fourier transforms, and/or correlation functions can also be used to characterize the dendrite arm spacing distribution.  However, it should be noted that these approaches are not intended for local characterization of the dendrite arm spacing and are not as effective for correlating the local spacing with local microstructure features as shown herein.  Moreover, these techniques do not quantify the number of nearest neighbors and are often coupled with Voronoi polygons to compute the nearest neighbor distributions.  Rather, these analysis methods are more effective at characterizing and comparing the homogeneity/heterogeneity of the dendritic structure between different processing conditions.  Hence, there will be limited discussion of these techniques in the present work.

\section{Results}

\subsection{Application to dendritic microstructure}

\begin{figure}[bht!]
\centering
\includegraphics[width = \textwidth]{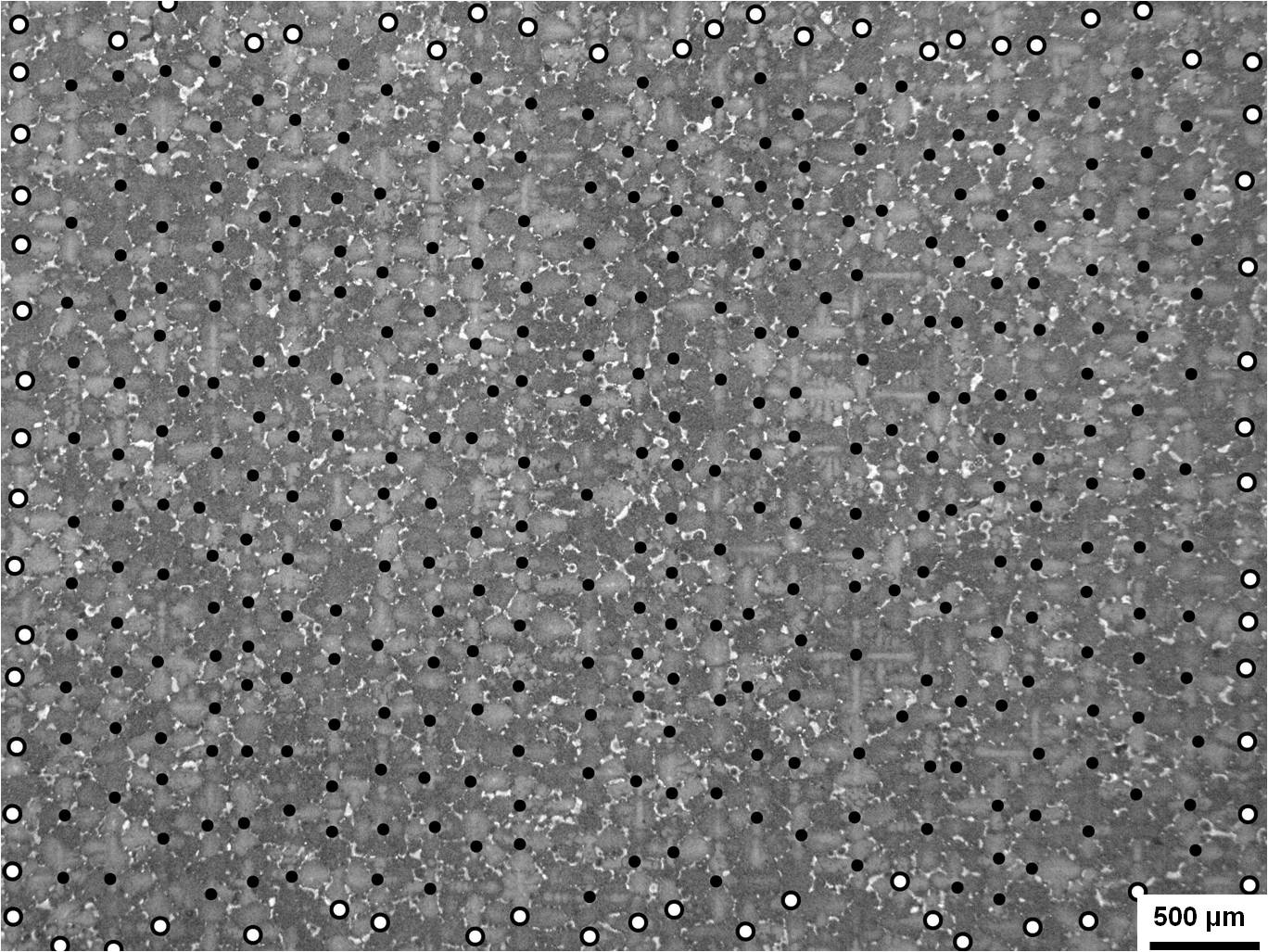}
\caption[]{Dendritic structure normal to the withdrawal direction in a directionally-solidified single crystal nickel-based superalloy cast using the liquid metal cooling technique \citep{Mil2011}.  The dots denote the dendrite cores, where the white dots indicate `edge' dendrite cores, as discussed in Figure~\ref{sx_nickel2} and the associated text. }
\label{sx_nickel}
\end{figure}

A micrograph of a directionally-solidified single crystal nickel-based superalloy microstructure that is polished and imaged perpendicular to the solidification direction is shown in Figure~\ref{sx_nickel}. This microstructure was produced using the liquid metal cooling technique, as described in Miller \citep{Mil2011} and Elliott et al.~\citep{Ell2004}. First, the dendrite cores were identified manually (white and black dots). Automated methods to identify dendrite cores can be invaluable for future large scale analysis \citep{Tsc2010a,Tsc2010b}. Moreover, the white particles in this image are eutectic particles. A total of 393 dendrite cores are contained in this image over an area of 24.25 mm$^2$, giving a PDAS of 248.4 $\mu$m using $c=1$ (Equation~\ref{lambda}). The remainder of the analysis uses this micrograph as a template for characterizing the local dendrite arm spacing.

\subsection{Accounting for image/part edge effects}

\begin{figure}[bht!]
\centering
        \begin{subfigure}[b]{0.75\textwidth}
                \includegraphics[width=\textwidth]{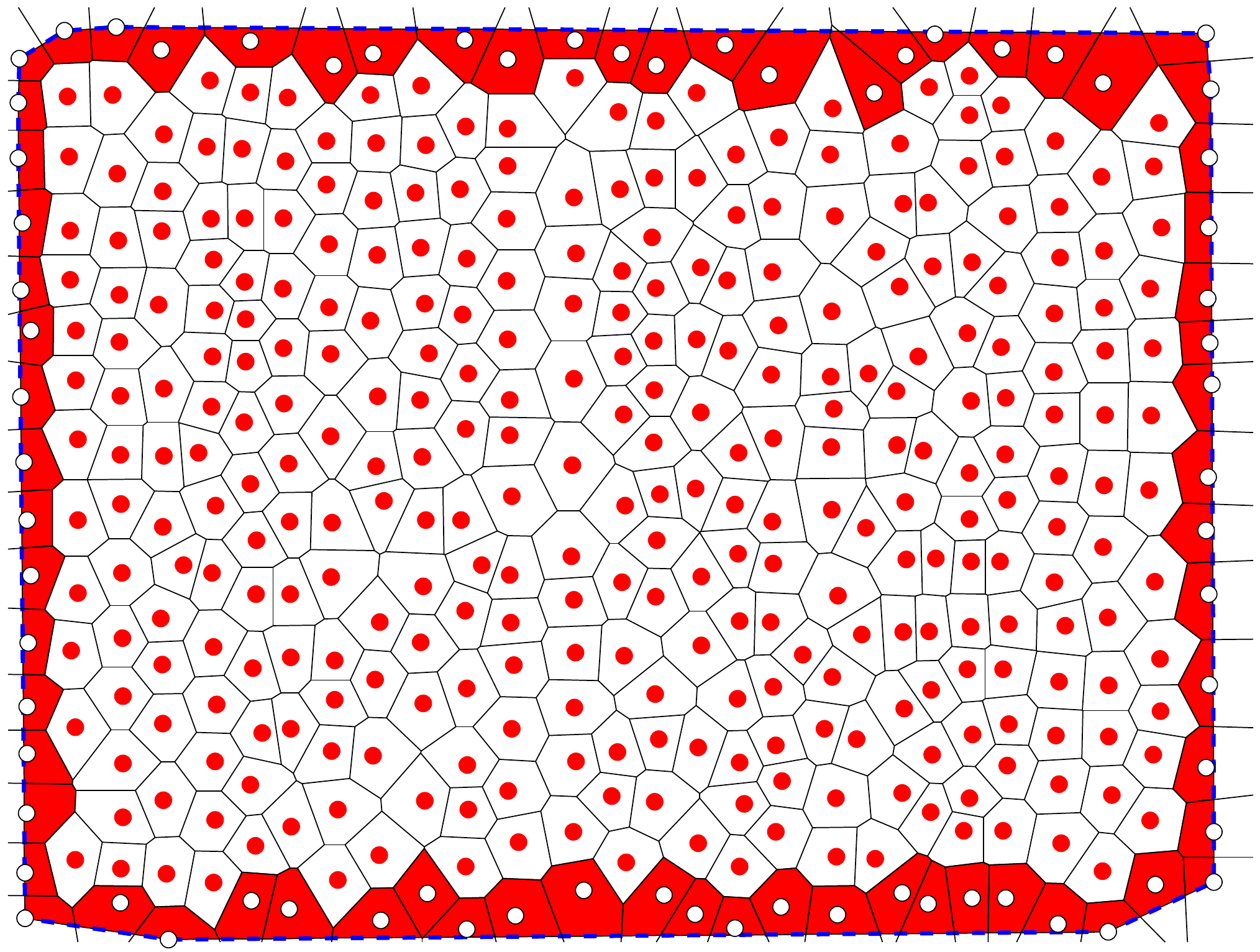}
                \caption{Convex hull}
                \label{hulla}
        \end{subfigure}%

        \begin{subfigure}[b]{0.475\textwidth}
                \includegraphics[width=\textwidth]{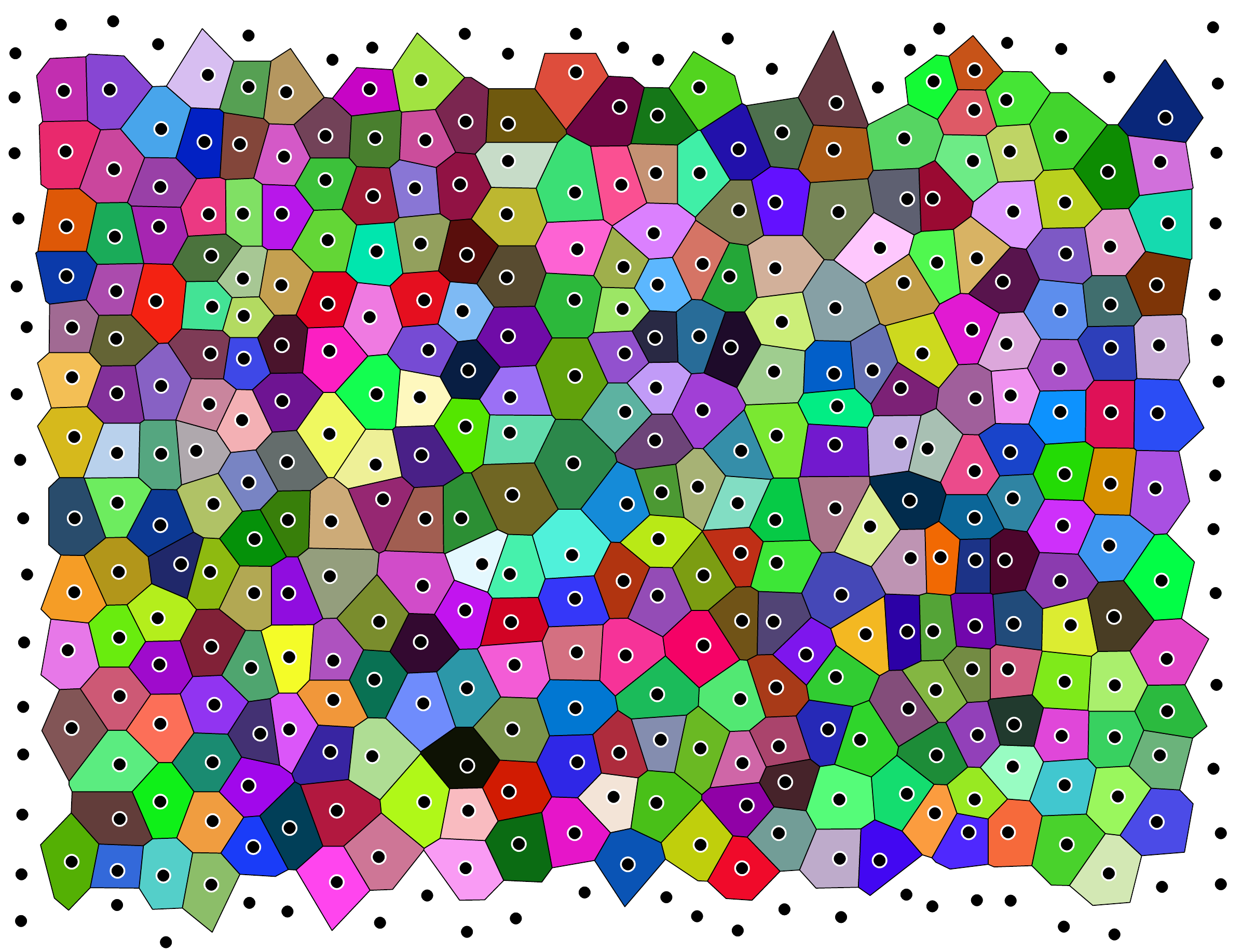}
                \caption{Interior dendrite cores}
                \label{hullb}
        \end{subfigure}%
        \begin{subfigure}[b]{0.475\textwidth}
                \includegraphics[width=\textwidth]{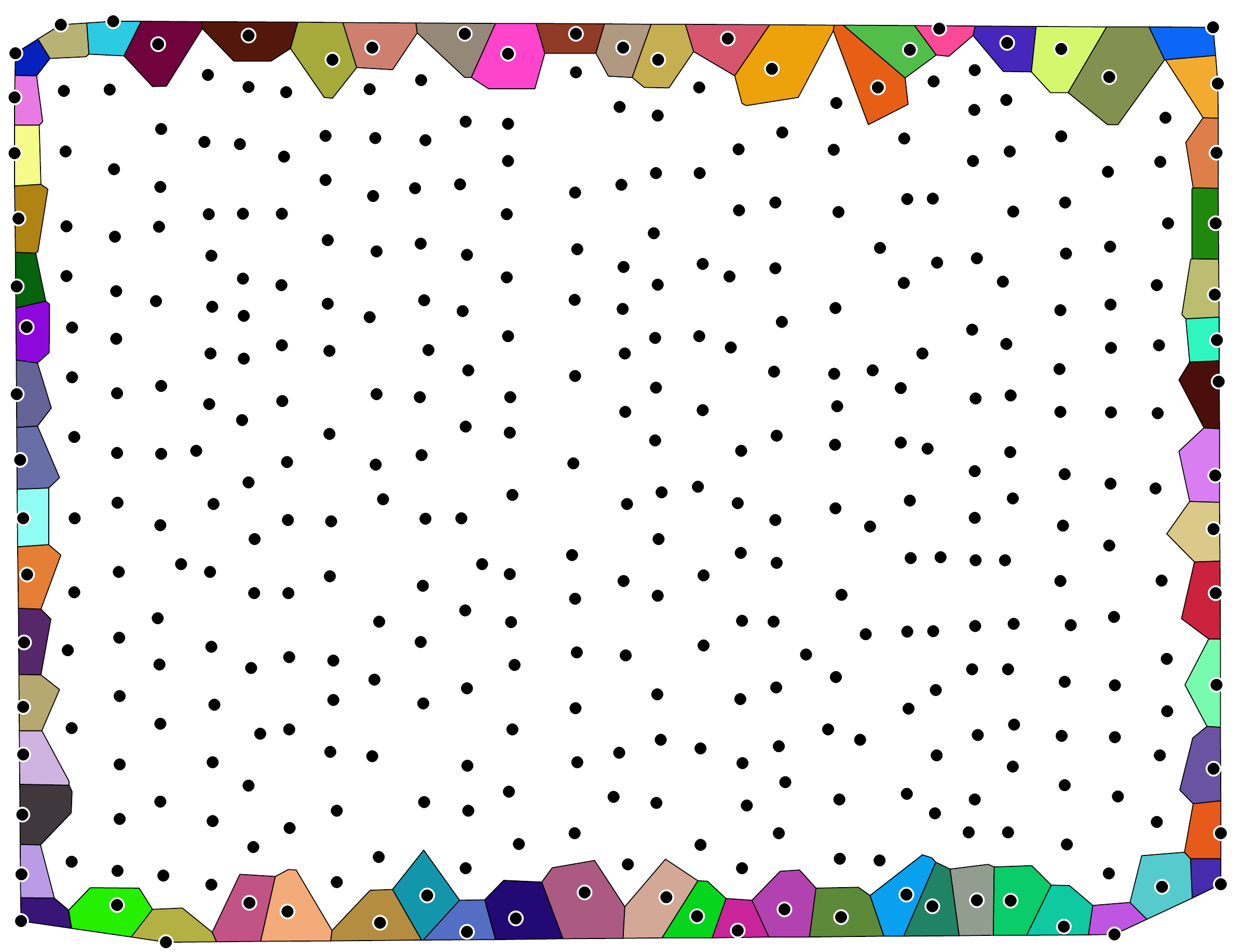}
                \caption{Edge dendrite cores}
                \label{hullc}
        \end{subfigure}
\caption[]{(a) Voronoi tessellation of dendritic structure from Figure \ref{sx_nickel}.  The dotted blue line (surrounding the points) denotes the convex hull of the dendrite cores and the red polygons delineate the cores that intersect the convex hull.  The interior and edge dendrites are shown in (b) and (c), respectively, with each polygon colored differently as a guide to the eye.}
\label{sx_nickel2}
\end{figure}

The ability to handle edge effects when computing local dendrite arm spacings with dendrite cores is vital for quantifying statistics in thin sections, such as the wall of an airfoil blade that may only contain 1--3 dendrite cores across the section \citep[e.g.,][]{Tsc2010a,Tsc2010b}. As a first example of one such a technique, we have used a convex hull of the dendrite cores in Figure~\ref{sx_nickel} to identify ``edge'' dendrite cores and quantify the dendrite arm spacing.  The dendrite core locations are first extracted from the experimental image, as shown in Figure~\ref{sx_nickel2}.  Then, a convex hull is generated around the points; this is the minimum ``convex'' area that contains all the points. Next, the edge points (white dots in Fig.~\ref{sx_nickel}) are identified by finding those points with Voronoi vertices that lie outside of the convex hull (dotted blue line in Figure~\ref{hulla}). Then, to utilize Voronoi-based techniques for these points, a new polygon is generated by the intersection of the initial polygon from the Voronoi tessellation and the convex hull; the new polygon of the edge dendrite cores is colored red in \ref{hulla} to distinguish from the bulk dendrite cores.  The polygons belonging to the interior and edge dendrites are shown in Figures \ref{hullb} and \ref{hullc}, with a random coloring scheme used to delineate the different polygons.  Last, the neighbors can now be calculated using either a new criterion or the same criterion used for interior points. For the present analysis, the same criterion (polygon with edge length threshold) was used for all points; although herein the interior dendrite cores are used to compare statistics with other techniques and bulk PDAS values. More complicated techniques are needed to deal with complex geometries that include concave character and internal passages in order to eventually apply these techniques to complex structures such as turbine blades.  Multiple instantiations of microstructures with edge effects can shed light on the appropriate method for determining the local PDAS at edges, which may be different from that used in the interior.

\subsection{Spatial distribution of local primary dendrite arm spacings}

\begin{figure}[bht!]
        \centering
        \begin{subfigure}[b]{0.45\textwidth}
                \centering
                \includegraphics[trim=0 225 0 0, clip,width=\textwidth]{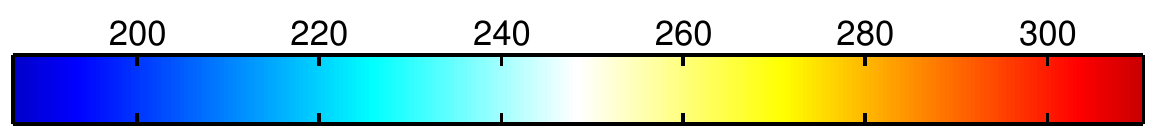}
                \phantomsubcaption
        \end{subfigure}%
\quad
        \begin{subfigure}[b]{0.45\textwidth}
                \centering
                \includegraphics[trim=0 225 0 0, clip,width=\textwidth]{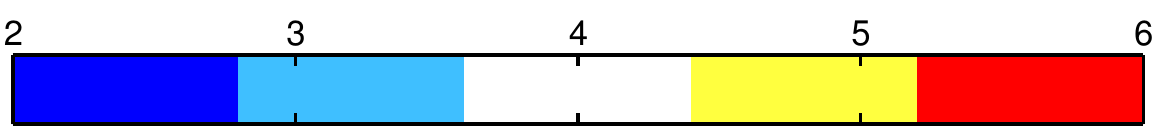}
                \phantomsubcaption
        \end{subfigure}
    \setcounter{subfigure}{0}
\\
        \begin{subfigure}[b]{0.45\textwidth}
                \centering
                \includegraphics[width=\textwidth]{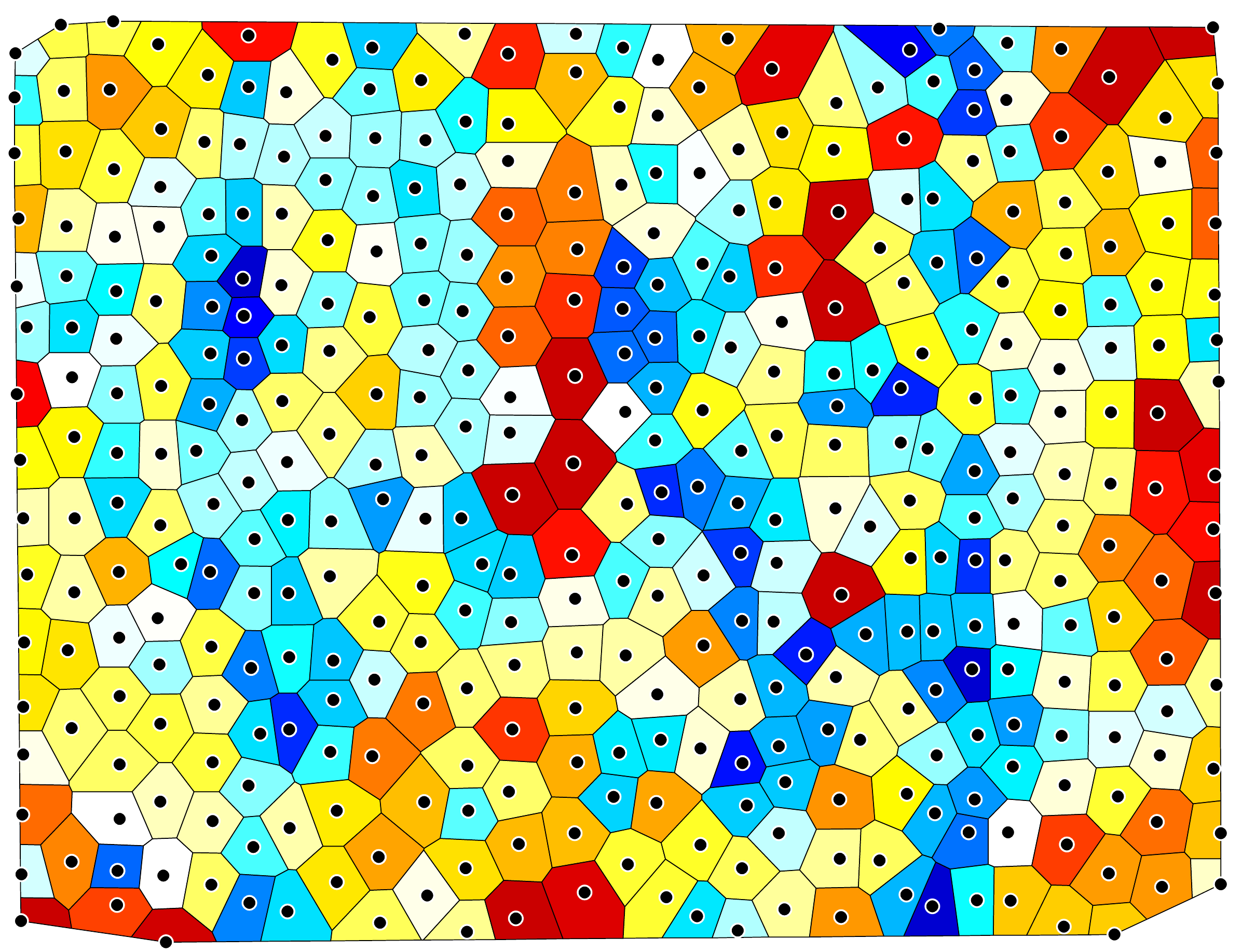}
                \caption{Primary dendrite arm spacing ($\mu$m)}
        \end{subfigure}
\quad
        \begin{subfigure}[b]{0.45\textwidth}
                \centering
                \includegraphics[width=\textwidth]{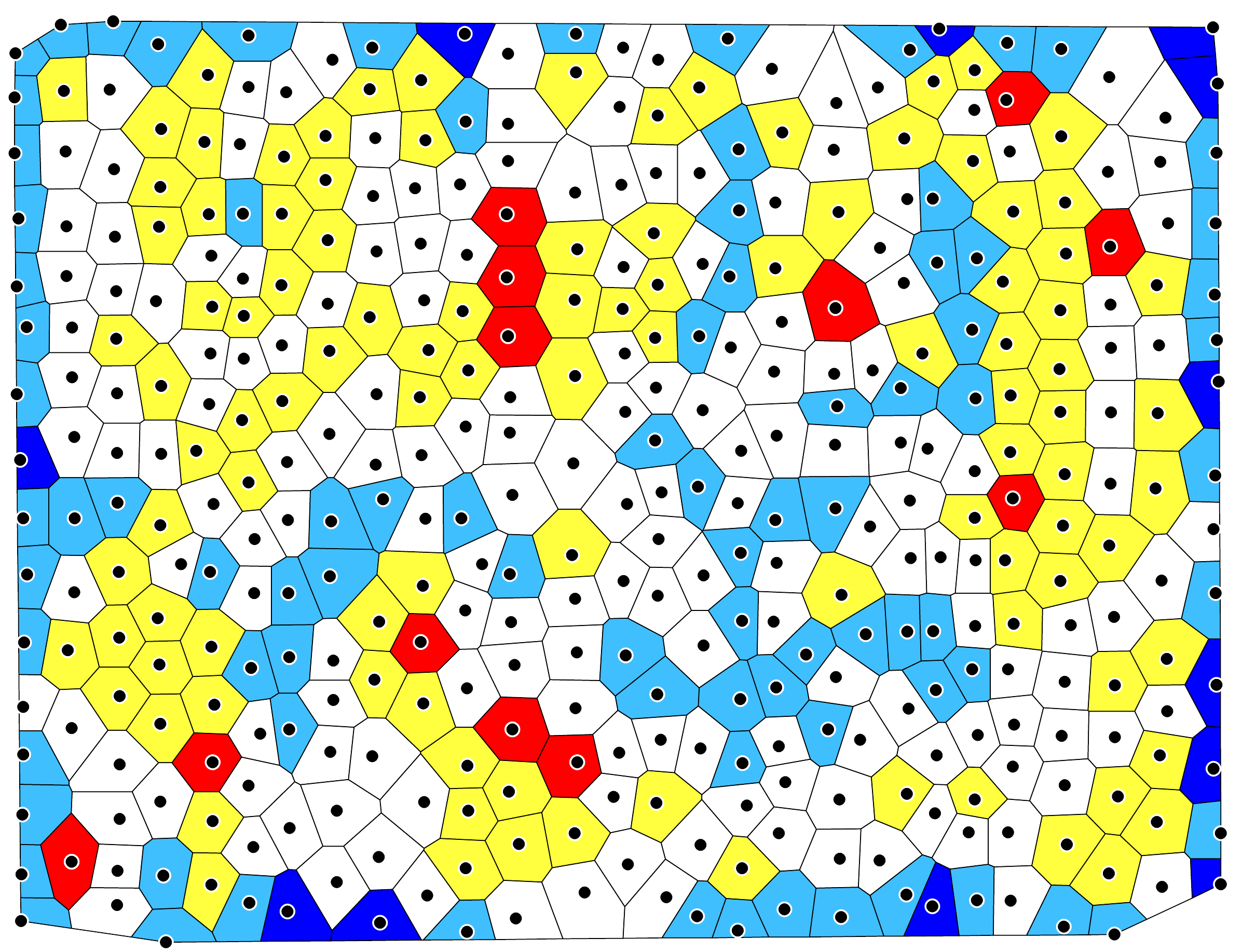}
                \caption{Coordination number}
        \end{subfigure}
        \caption{(a) Local dendrite arm spacing ($\mu$m) and (b) coordination number based on the Voronoi tessellation with edge length threshold of $d_{crit}$=0.12 or 12\%.}
\label{dendrite}
\end{figure}

The spatial distribution of local dendrite arm spacing and coordination number can provide insight into the order/disorder of primary dendrites and can identify regions that could potentially contain more/less interdendritic features and/or contain different properties. For instance, the primary dendrite arm spacing and coordination number for the directionally-solidified superalloy micrograph (Figure~\ref{sx_nickel}) is shown in Figure~\ref{dendrite}. In this example, we used the Voronoi tessellation-based technique with an edge length threshold of $d_{crit} = 0.12$. Dendrite cores with local PDAS similar to the mean PDAS of the bulk (248.4 $\mu$m) are colored white and those with PDAS above (below) the mean PDAS are red (blue); the lower and upper bounds of the colorbar are -25\% and +25\% of the mean PDAS value, respectively. In general, the exterior dendrite cores have similar PDAS as the interior dendrite cores using this technique. A similar colorbar is used for the coordination number as well. As would be expected, the exterior dendrite cores tend to have a lower coordination number than the interior dendrite cores, with a few that only have 2 nearest neighbors. However, the dendrite cores with a low coordination number on the edges are not consistently over/under the mean PDAS (i.e., they do not significantly bias the statistics from the edge dendrite cores).  Future work will examine what techniques may be most applicable for characterizing local dendrite arm spacings and coordination numbers for dendrite cores on free surfaces.  It is envisioned that sectioning large numbers of instantiations of synthetically-generated microstructures of known bulk dendrite arm spacings can be used to understand the bias introduced by edge effects and to understand what are the best techniques for quantifying the local spacing.

\begin{figure}[bht!]
        \centering
        \begin{subfigure}[b]{0.75\textwidth}
                \centering
                \includegraphics[trim=0 225 0 0, clip,width=\textwidth]{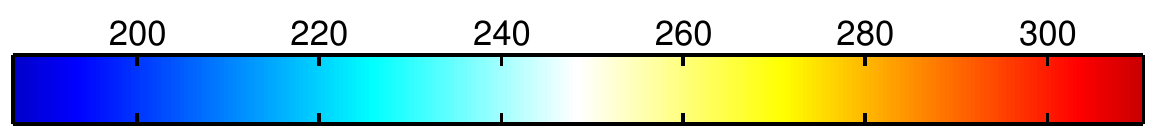}
                \phantomsubcaption
        \end{subfigure}%
    \setcounter{subfigure}{0}
\\
        \begin{subfigure}[b]{0.45\textwidth}
                \centering
                \includegraphics[width=\textwidth]{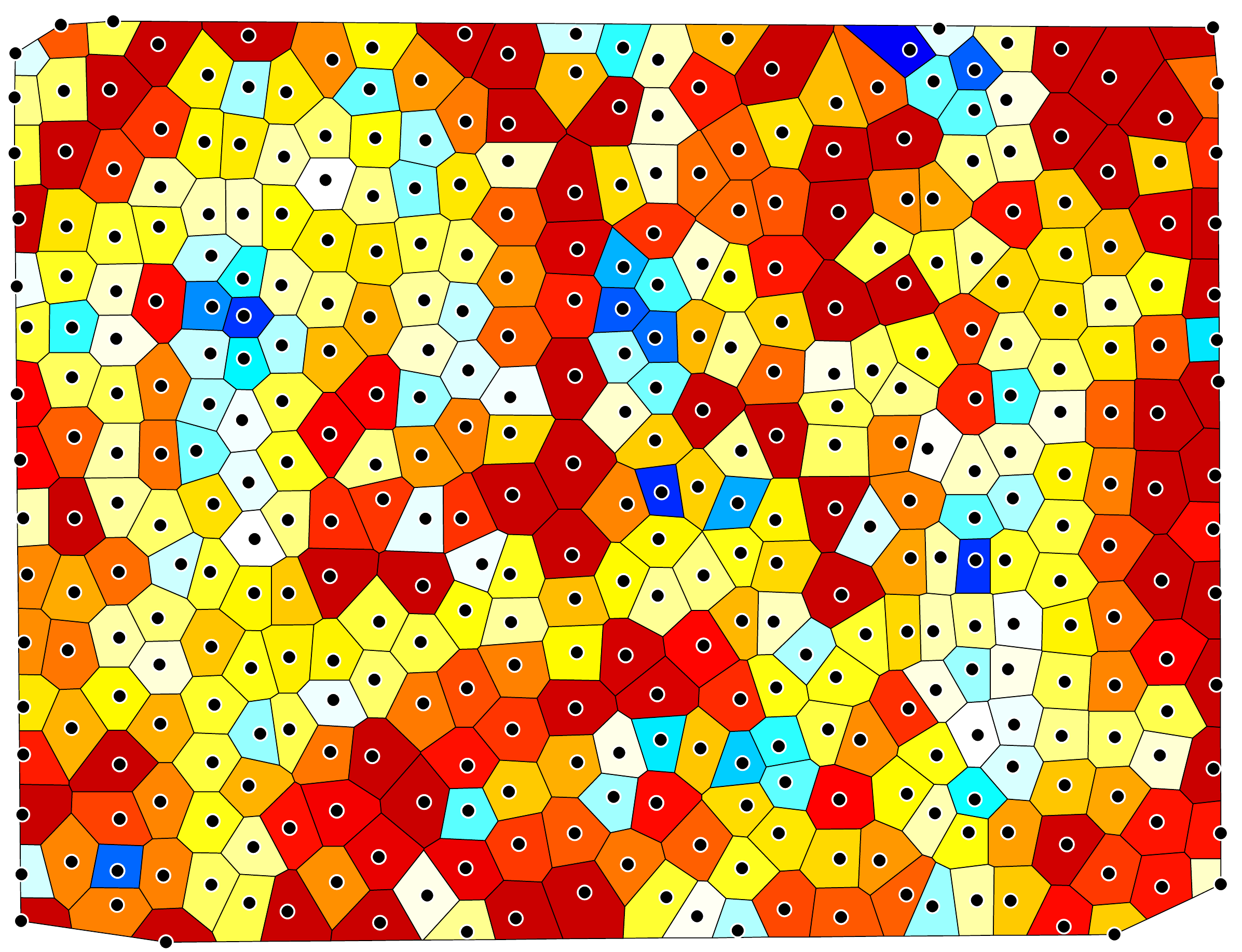}
                \caption{Voronoi Tesselation ($d_{crit}=0.0$)}
        \end{subfigure}
\quad
        \begin{subfigure}[b]{0.45\textwidth}
                \centering
                \includegraphics[width=\textwidth]{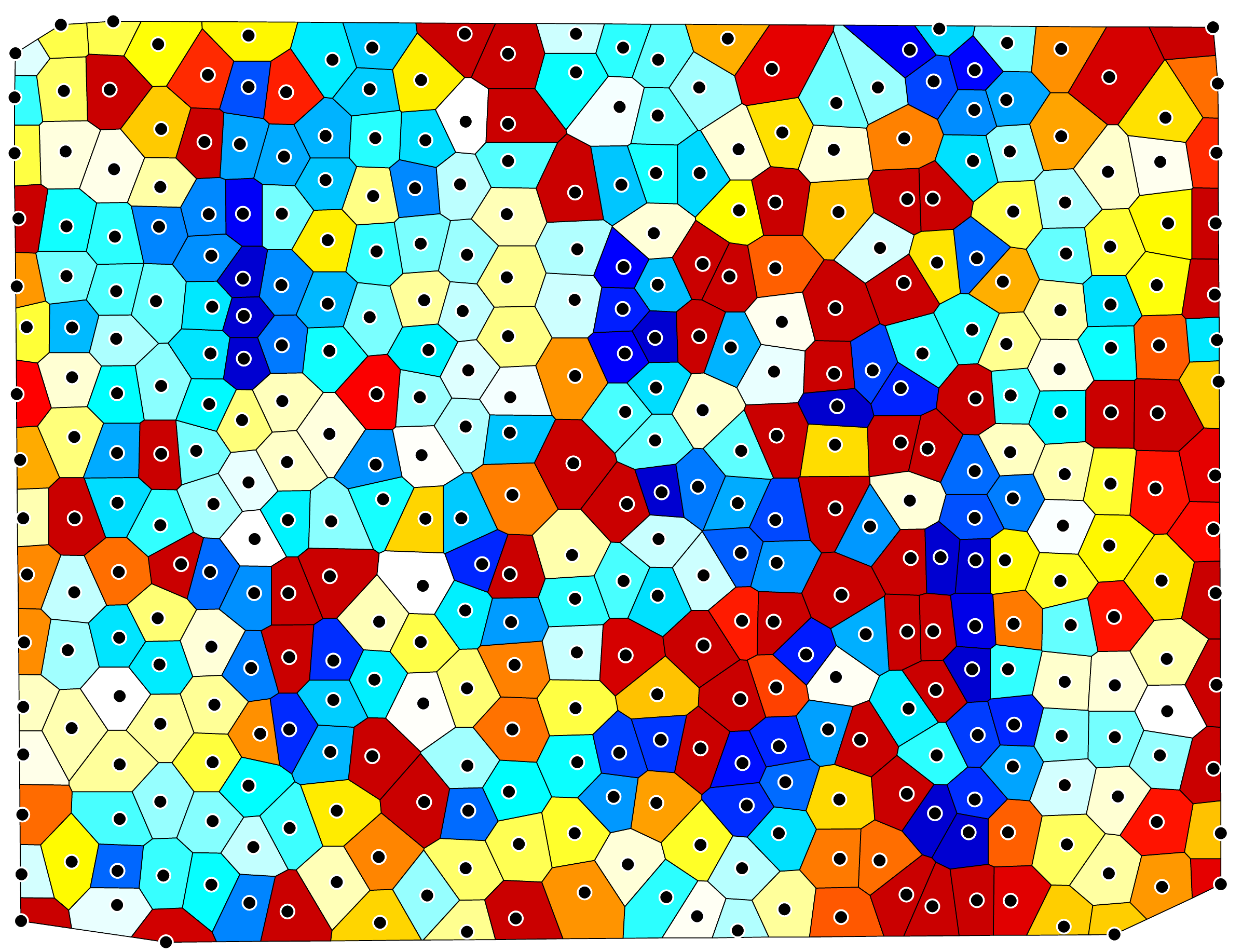}
                \caption{Warnken-Reed ($\alpha=2.0$)}
                \label{wr}
        \end{subfigure}
\quad
        \begin{subfigure}[b]{0.45\textwidth}
                \centering
                \includegraphics[width=\textwidth]{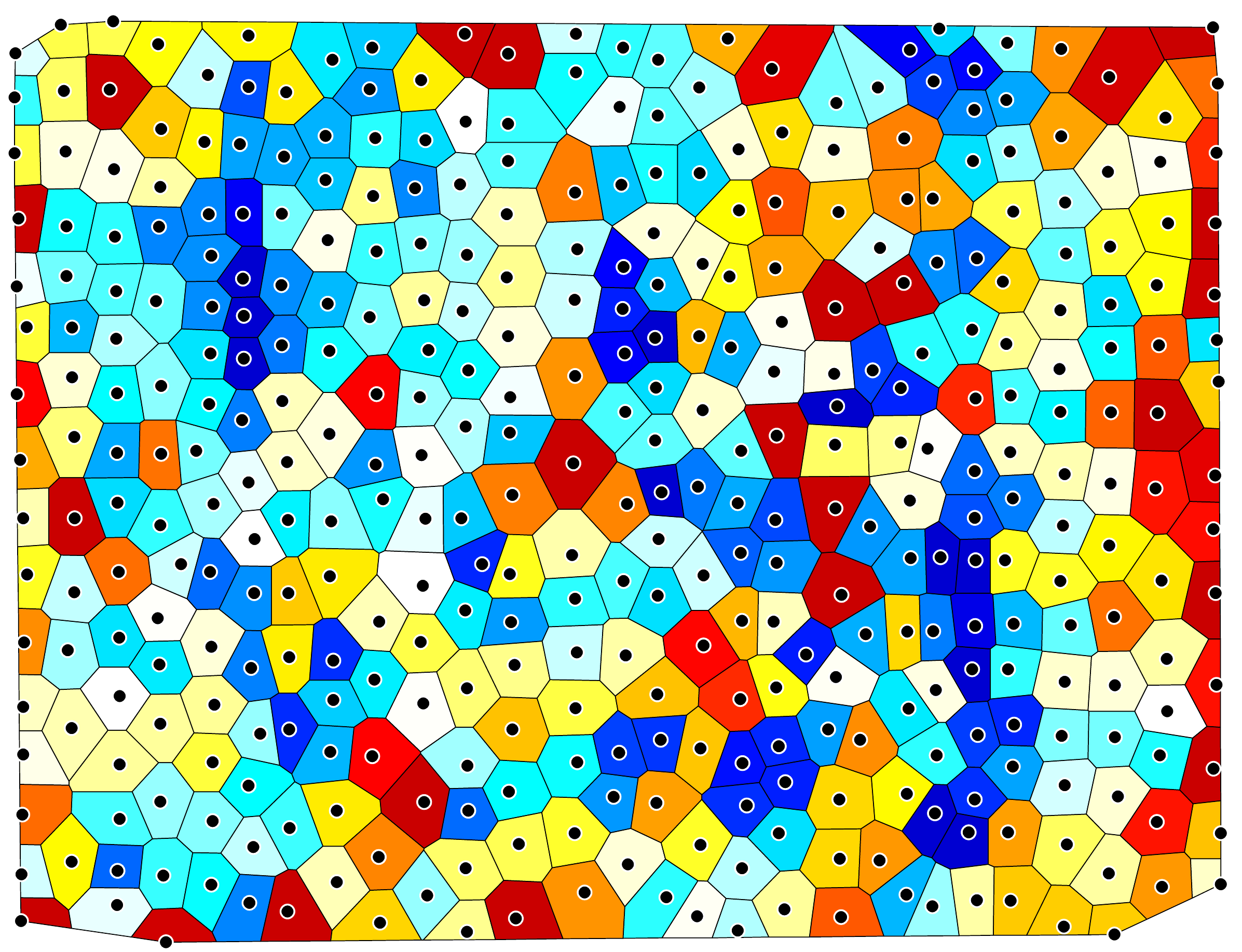}
                \caption{Voronoi Warnken-Reed ($\alpha=2.0$)}
                \label{mwr}
        \end{subfigure}
        \caption{Local dendrite arm spacing ($\mu$m) for the three techniques not shown in Figure~\ref{dendrite}: (a) Voronoi tessellation with edge length threshold of $d_{crit}$=0.0, (b) Warnken--Reed technique with $\alpha=2.0$, and (c) Voronoi Warnken-Reed with $\alpha=2.0$.}
\label{dendrite2}
\end{figure}

The local dendrite arm spacing for the remaining three techniques is shown in Figure~\ref{dendrite2}.  The same color bar for local PDAS as in Figure~\ref{dendrite} is used here.  First, notice that the Voronoi tessellation-based technique with an edge length threshold of $d_{crit} = 0.0$ has a much larger fraction of dendrite cores with PDAS greater than the bulk PDAS than below the bulk PDAS (83.5\% above 248.4 $\mu$m).  Clearly, the local primary dendrite arm spacing is overpredicted in this case.  The Warnken--Reed and Voronoi Warnken--Reed methods are shown in Figures~\ref{wr} and \ref{mwr}.  At first glance, a majority of the local PDAS values are very similar between the two methods ($\sim$79\%).  However, $\sim$21\% of the cores resulted in a difference between the two techniques, which is caused by the original Warnken--Reed method using neighbors outside of those FNNs identified from the Voronoi polygons.  In every case, the Warnken--Reed method resulted in higher local PDAS values than the Voronoi Warnken--Reed method, as would be expected since this is purely a distance-based method and subsequent additions can only increase the local PDAS.  

\begin{figure}[bht!]
        \centering
        \begin{subfigure}[b]{0.75\textwidth}
                \centering
                \includegraphics[trim=0 225 0 0, clip,width=\textwidth]{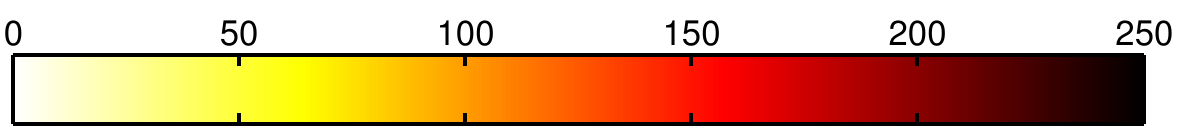}
                \phantomsubcaption
        \end{subfigure}%
    \setcounter{subfigure}{0}
\\
        \begin{subfigure}[b]{0.75\textwidth}
                \centering
                \includegraphics[width=\textwidth]{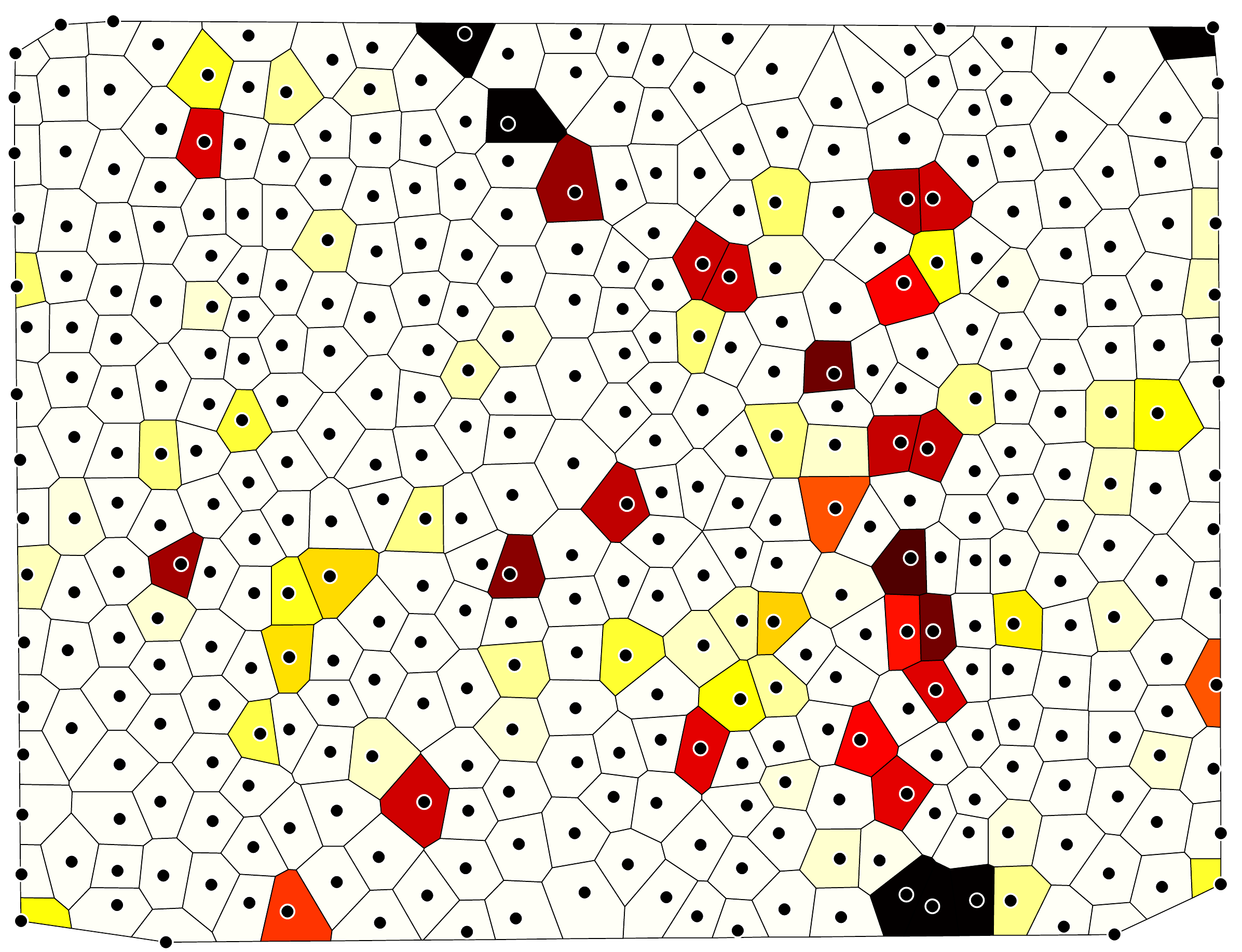}
                \caption{Primary dendrite arm spacing ($\mu$m) difference}
        \end{subfigure}
        \caption{Difference in the local dendrite arm spacing ($\mu$m) between the Warnken--Reed and Voronoi Warnken--Reed techniques with $\alpha=2.0$.  The Warnken--Reed technique had a greater PDAS value in every case ($\sim$21\% of dendrite cores are different).}
\label{dendrite3}
\end{figure}

Figure \ref{dendrite3} shows the difference in local PDAS values between the two techniques.  In several cases, the difference is greater than 250 $\mu$m and/or 100\% of the PDAS value quantified by the Voronoi Warnken-Reed method.  The differing dendrite cores is approximately an equal percentage for edge dendrites as well as interior dendrite cores.  For some cases, it is apparent that one of the closest three dendrite cores is significantly closer or further away than the other two, thereby resulting in a larger standard deviation $d_{std}$ and a greater chance to add multiple neighbors; this case is similar to that shown in Figure \ref{various_methods2}.

\subsection{Local primary dendrite arm spacing statistics}

\begin{figure}[bht!]
        \centering
        \begin{subfigure}[b]{0.45\textwidth}
                \centering
                \includegraphics[width=\textwidth]{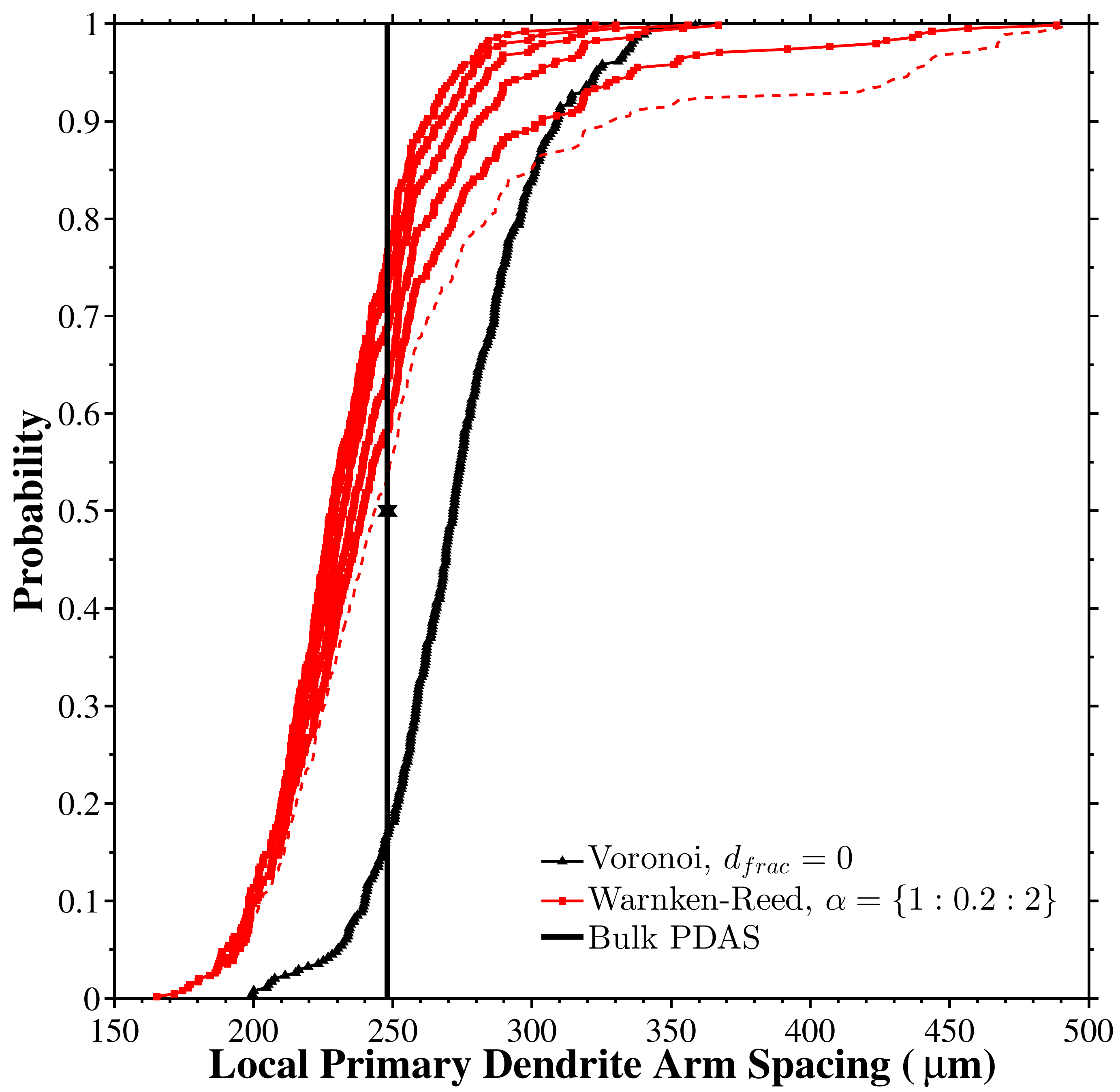}
                \caption{}
        \end{subfigure}%
\quad
        \begin{subfigure}[b]{0.45\textwidth}
                \centering
                \includegraphics[width=\textwidth]{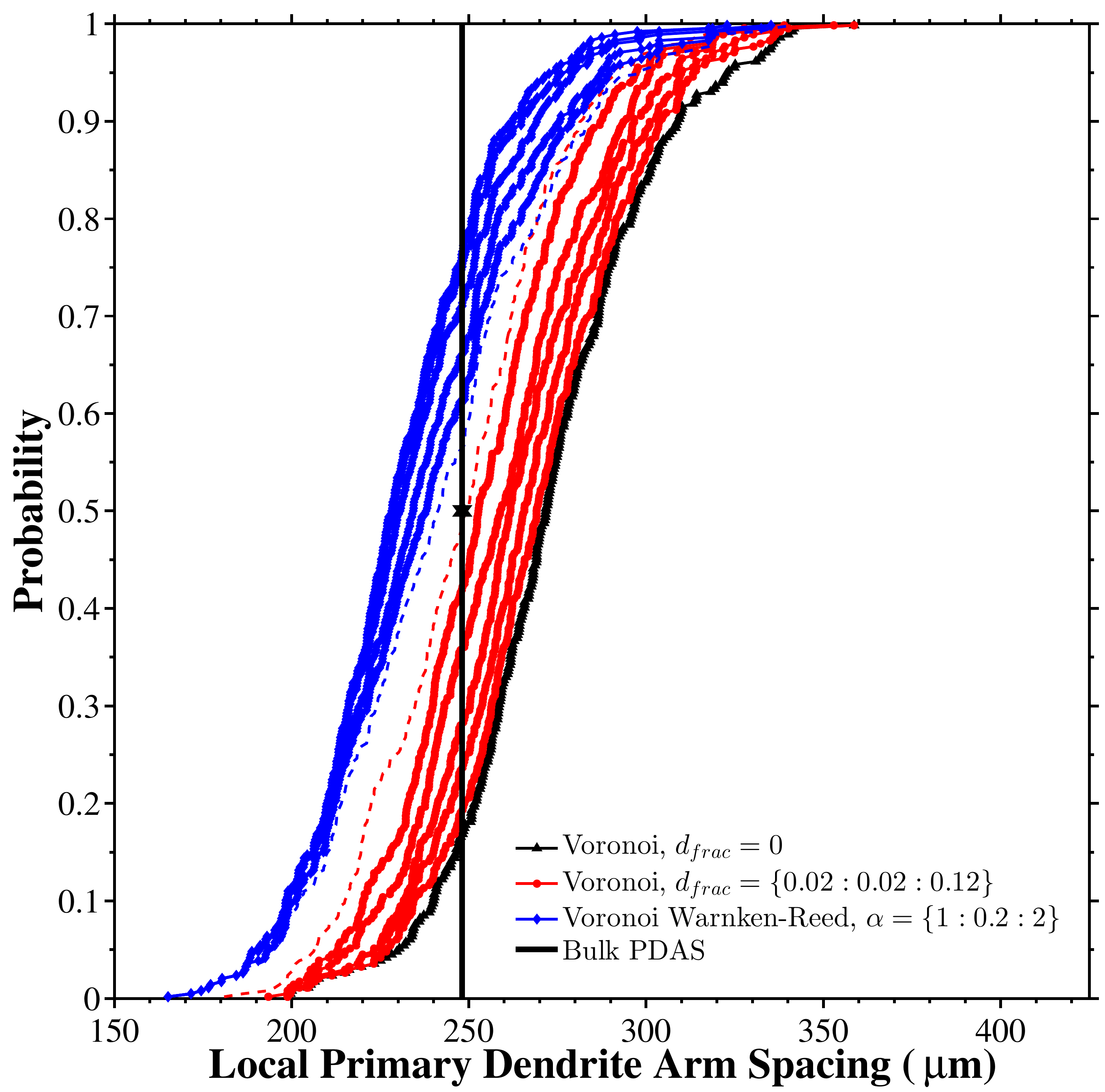}
                \caption{}
        \end{subfigure}

\caption[]{Probability distribution functions for the various local characterization methods compared within for the internal dendrites within the dendritic microstructure shown in Figure~\ref{sx_nickel}. The four different techniques are compared with the bulk PDAS for a range of parameter values.  The upper bound of the parameter range for each technique is shown as a dotted line.  To facilitate the comparison, the Warnken--Reed and the Voronoi technique ($d_{crit}=0.0$) are shown in (a), and the remaining Voronoi-modified techniques are shown in (b).}
\label{probability}
\end{figure}

The local dendrite arm spacing statistics are also calculated for the interior dendrite cores to compare with the traditional PDAS measurement. The cumulative distribution function plot for the local dendrite arm spacing is shown in Figure~\ref{probability} for the three different techniques over a range of parameter values, which are given in the legend. The bulk PDAS measurement is shown as a vertical black line and the hexagonal star shows the 50$^\textrm{th}$ percentile intersection point. The local dendrite arm spacing distributions are characterized in terms of mean, standard deviation, skewness, and kurtosis (Table \ref{table1}), while the coordination number distributions are characterized in terms of their mean and the percentages of 3, 4, 5, 6, and 7+ nearest neighbors (Table \ref{table2}). The skewness and kurtosis measure the asymmetry of the distribution and the sharpness of the peak/thickness of the tail, respectively. The skewness and kurtosis are 0 and 3, respectively, for a normal distribution.

There are distinct differences between the local dendrite arm spacing distributions calculated by the four techniques (Figure~\ref{probability}, Tables \ref{table1}). The Warnken--Reed and Voronoi Warnken--Reed are compared initially. In the case of the Warnken--Reed method, the PDAS distribution is shifted towards large PDAS values at high $\alpha$ values (a positive skewness value gives a long tail) and has a sharper peak and a longer, fatter tail (high kurtosis values), more so than the other methods. This skewness is caused by an overestimation of the number of nearest neighbors in some cases, due to large values of either $d_{std}$ or the parameter $\alpha$. Hence, while the calculated mean PDAS can approach the bulk-measured value of 248.4 $\mu$m (within 0.1\% for $\alpha$=1.8), this mean PDAS is highly sensitive to these large PDAS values. This overprediction of nearest neighbors, and their result on the local PDAS distribution, is also apparent by comparing this with the Voronoi Warnken--Reed method, whereby the potential nearest neighbors are restricted to only those FNNs defined by the Voronoi polygon. In this case, there is a lack of a long tail and the skewness/kurtosis of the distribution tends more towards normality. However, the calculated mean PDAS with this method tends to underestimate the bulk-measured PDAS. While the maximum number of nearest neighbors (8 for $\alpha \ge 1.2$) is more realistic, a large percentage of dendrite cores are predicted to have only 3 nearest neighbors, even in the case of a large $\alpha$ parameter (48.6\% for $\alpha =2.0$).  It is also interesting that increasing the $\alpha$ parameter for the case of the Voronoi Warnken--Reed method tends to shift the slope of the probability distribution function without affecting either the minimum or maximum local dendrite arm spacings. 

For comparison, the minimal spanning tree method (Fig.~\ref{MST}) was also included in Table \ref{table1}.  Not surprisingly, the mean distance of the connecting line segments is much shorter than the bulk calculated PDAS using Eq.~\ref{lambda} with $c=1$.  Remember that the MST method is composed of the shortest line segments to connect all dendrites.  Both the MST standard deviation and kurtosis values are larger than the Voronoi tesselation method (for all $d_{crit}$) and the Voronoi Warnken-Reed method (for all $\alpha$), indicating a wider distribution and a larger deviation of the distribution from normality (kurtosis = 3).  Moreover, the distribution is skewed towards a larger tail at lower distances (negative skewness) unlike the other techniques, which again is associated with the selection of the shortest line segments to characterize the spacing. 

\begin{table}[ht]
  \centering
\footnotesize
  \caption{Local primary dendrite arm spacing statistics}
    \begin{tabular}{c c c . c cc}
    \addlinespace
    \toprule
\multicolumn{1}{c}{} &
\multicolumn{1}{c}{} &
\multicolumn{5}{c}{Primary Dendrite Arm Spacing} \\

\multicolumn{1}{c}{Method} &
\multicolumn{1}{c}{Parameter} &
\multicolumn{1}{c}{mean, $\mu$m} &
\multicolumn{1}{c}{diff, \%} &
\multicolumn{1}{c}{std, $\mu$m} &
\multicolumn{1}{c}{skewness} &
\multicolumn{1}{c}{kurtosis} \\

    \midrule
	Bulk (Eq.~\ref{lambda}, $c=1$) & - & 248.4 & 0.0 & - & - & - \\
    Voronoi Tesselation & $d_{crit}$ = 0.00 & 272.9 & 9.9 & 28.0  & 0.1   & 3.3 \\
    Voronoi Tesselation & $d_{crit}$ = 0.02 & 270.0 & 8.7 & 27.2  & 0.1   & 3.2 \\
    Voronoi Tesselation & $d_{crit}$ = 0.04 & 266.4 & 7.2 & 26.8  & 0.1   & 3.1 \\
    Voronoi Tesselation & $d_{crit}$ = 0.06 & 263.0 & 5.9 & 25.7  & 0.1   & 3.1 \\
    Voronoi Tesselation & $d_{crit}$ = 0.08 & 258.4 & 4.0 & 26.1  & 0.2   & 3.1 \\
    Voronoi Tesselation & $d_{crit}$ = 0.10 & 253.3 & 2.0 & 25.3  & 0.2   & 3.0 \\
    Voronoi Tesselation & $d_{crit}$ = 0.12 & 247.6 & -0.3 & 26.0  & 0.2   & 2.9 \\
    Voronoi Warnken--Reed & \ensuremath{\alpha} = 1.0 & 230.0 & -7.4 & 25.0  & 0.4   & 3.5 \\
    Voronoi Warnken--Reed & \ensuremath{\alpha} = 1.2 & 230.9 & -7.0 & 26.0  & 0.4   & 3.4 \\
    Voronoi Warnken--Reed & \ensuremath{\alpha} = 1.4 & 232.7 & -6.3 & 27.4  & 0.4   & 3.2 \\
    Voronoi Warnken--Reed & \ensuremath{\alpha} = 1.6 & 236.1 & -5.0 & 29.5  & 0.4   & 3.1 \\
    Voronoi Warnken--Reed & \ensuremath{\alpha} = 1.8 & 239.0 & -3.8 & 30.5  & 0.4   & 3.0 \\
    Voronoi Warnken--Reed & \ensuremath{\alpha} = 2.0 & 242.2 & -2.5 & 31.8  & 0.3   & 3.0 \\
    Warnken--Reed & \ensuremath{\alpha} = 1.0 & 230.1 & -7.4 & 25.1  & 0.3   & 3.4 \\
    Warnken--Reed & \ensuremath{\alpha} = 1.2 & 231.8 & -6.7 & 26.9  & 0.5   & 3.5 \\
    Warnken--Reed & \ensuremath{\alpha} = 1.4 & 234.5 & -5.6 & 29.4  & 0.6   & 3.8 \\
    Warnken--Reed & \ensuremath{\alpha} = 1.6 & 239.2 & -3.7 & 33.0  & 0.8   & 4.0 \\
    Warnken--Reed & \ensuremath{\alpha} = 1.8 & 248.1 & -0.1 & 48.1  & 1.9   & 8.3 \\
    Warnken--Reed & \ensuremath{\alpha} = 2.0 & 259.2 & 4.3 & 64.2  & 1.9   & 6.5 \\
    Minimal spanning tree & N/A & 215.2 & -13.4 & 34.1 & -0.5 & 4.8 \\
    \bottomrule
    \end{tabular}%
  \label{table1}%
\end{table}%

The Voronoi tessellation techniques are also compared. First, quantifying the coordination number and the local PDAS values using all FNNs identified by the Voronoi tessellation polygons ($d_{crit}=0$) clearly overestimates both measures; mean PDAS is $\sim$10\% off from the bulk-measured PDAS value and $\sim$20\% of dendrite cores have more than 6 nearest neighbors. As the edge length threshold parameter increases, less nearest neighbors are identified and the calculated mean PDAS approaches the bulk-measured PDAS value (within 0.3\% for $d_{crit}=0.12$). For $d_{crit}=0.12$, the majority of dendrite cores have 4 nearest neighbors ($>$50\%), followed by 5 nearest neighbors (26.9\%) and 3 nearest neighbors (17.3\%). Moreover, the local PDAS distribution has a low skewness value (0.2) and a kurtosis of 2.9, indicating an approximately normal distribution. In general, the Voronoi tessellation-based technique with an edge length threshold criterion of $d_{crit}=0.12$ tends to give the best agreement in terms of both bulk-measured PDAS and coordination number. Furthermore, this technique allows for calculating the local PDAS value and the local PDAS distribution, which may be important for assessing the homogeneity of dendrite growth and/or for identifying local regions where the local growth conditions/properties are different from the norm.

\begin{table}[ht]
  \centering
\footnotesize
  \caption{Local coordination number statistics}
    \begin{tabular}{c c . . . . . c}
    \toprule

\multicolumn{1}{c}{} &
\multicolumn{1}{c}{} &
\multicolumn{5}{c}{Coordination Number (\%)} &
\multicolumn{1}{c}{} \\

\multicolumn{1}{c}{Method} &
\multicolumn{1}{c}{Parameter} &
\multicolumn{1}{c}{3} &
\multicolumn{1}{c}{4} &
\multicolumn{1}{c}{5} &
\multicolumn{1}{c}{6} &
\multicolumn{1}{c}{\ensuremath{\ge} 7} &
\multicolumn{1}{c}{Mean} \\
    \midrule
    Voronoi Tesselation & $d_{crit}$ = 0.00 & 0.0   & 2.5   & 20.4  & 57.3  & 19.8  & 5.98 \\
    Voronoi Tesselation & $d_{crit}$ = 0.02 & 0.0   & 4.0   & 26.0  & 57.0  & 13.0  & 5.80 \\
    Voronoi Tesselation & $d_{crit}$ = 0.04 & 0.0   & 7.4   & 34.7  & 51.4  & 6.5   & 5.57 \\
    Voronoi Tesselation & $d_{crit}$ = 0.06 & 0.0   & 11.8  & 45.2  & 39.9  & 3.1   & 5.35 \\
    Voronoi Tesselation & $d_{crit}$ = 0.08 & 1.5   & 24.1  & 49.2  & 24.8  & 0.3   & 4.98 \\
    Voronoi Tesselation & $d_{crit}$ = 0.10 & 5.3   & 43.0  & 41.5  & 10.2  & 0.0   & 4.57 \\
    Voronoi Tesselation & $d_{crit}$ = 0.12 & 17.3  & 52.3  & 26.9  & 3.4   & 0.0   & 4.16 \\
    Voronoi Warnken--Reed & \ensuremath{\alpha} = 1.0 & 94.1  & 4.6   & 0.9   & 0.0   & 0.3   & 3.08 \\
    Voronoi Warnken--Reed & \ensuremath{\alpha} = 1.2 & 87.6  & 8.7   & 2.8   & 0.3   & 0.6   & 3.18 \\
    Voronoi Warnken--Reed & \ensuremath{\alpha} = 1.4 & 79.3  & 11.8  & 5.3   & 2.5   & 1.2   & 3.35 \\
    Voronoi Warnken--Reed & \ensuremath{\alpha} = 1.6 & 65.9  & 18.0  & 8.0   & 5.3   & 2.8   & 3.62 \\
    Voronoi Warnken--Reed & \ensuremath{\alpha} = 1.8 & 56.3  & 19.8  & 12.1  & 8.4   & 3.4   & 3.84 \\
    Voronoi Warnken--Reed & \ensuremath{\alpha} = 2.0 & 48.6  & 19.8  & 14.9  & 11.5  & 5.3   & 4.07 \\
    Warnken--Reed & \ensuremath{\alpha} = 1.0 & 91.6  & 7.1   & 0.9   & 0.0   & 0.3   & 3.10 \\
    Warnken--Reed & \ensuremath{\alpha} = 1.2 & 83.9  & 9.6   & 4.0   & 1.2   & 1.2   & 3.27 \\
    Warnken--Reed & \ensuremath{\alpha} = 1.4 & 72.8  & 14.6  & 7.1   & 2.8   & 2.8   & 3.52 \\
    Warnken--Reed & \ensuremath{\alpha} = 1.6 & 58.5  & 21.7  & 9.0   & 4.0   & 6.8   & 3.89 \\
    Warnken--Reed & \ensuremath{\alpha} = 1.8 & 52.0  & 19.2  & 10.8  & 6.2   & 11.8  & 4.63 \\
    Warnken--Reed & \ensuremath{\alpha} = 2.0 & 44.0  & 19.2  & 12.1  & 7.1   & 17.6  & 5.55 \\
    \bottomrule
    \end{tabular}%
  \label{table2}%
\end{table}%

\subsection{Correlation with interdendritic features}
The relationship between the occurrence of interdendritic features (e.g., pores or eutectic particles) and the local dendrite arm spacing (or distance from cores, etc.) can provide insight into the importance of quantifying the local spacings.  We have examined how these metrics may relate to the formation of eutectic particles in this work by first segmenting the interdendritic particles and then computing probability distribution functions.
	
\begin{figure}[bht!]
        \centering
        \begin{subfigure}[b]{0.31\textwidth}
                \includegraphics[width=\textwidth]{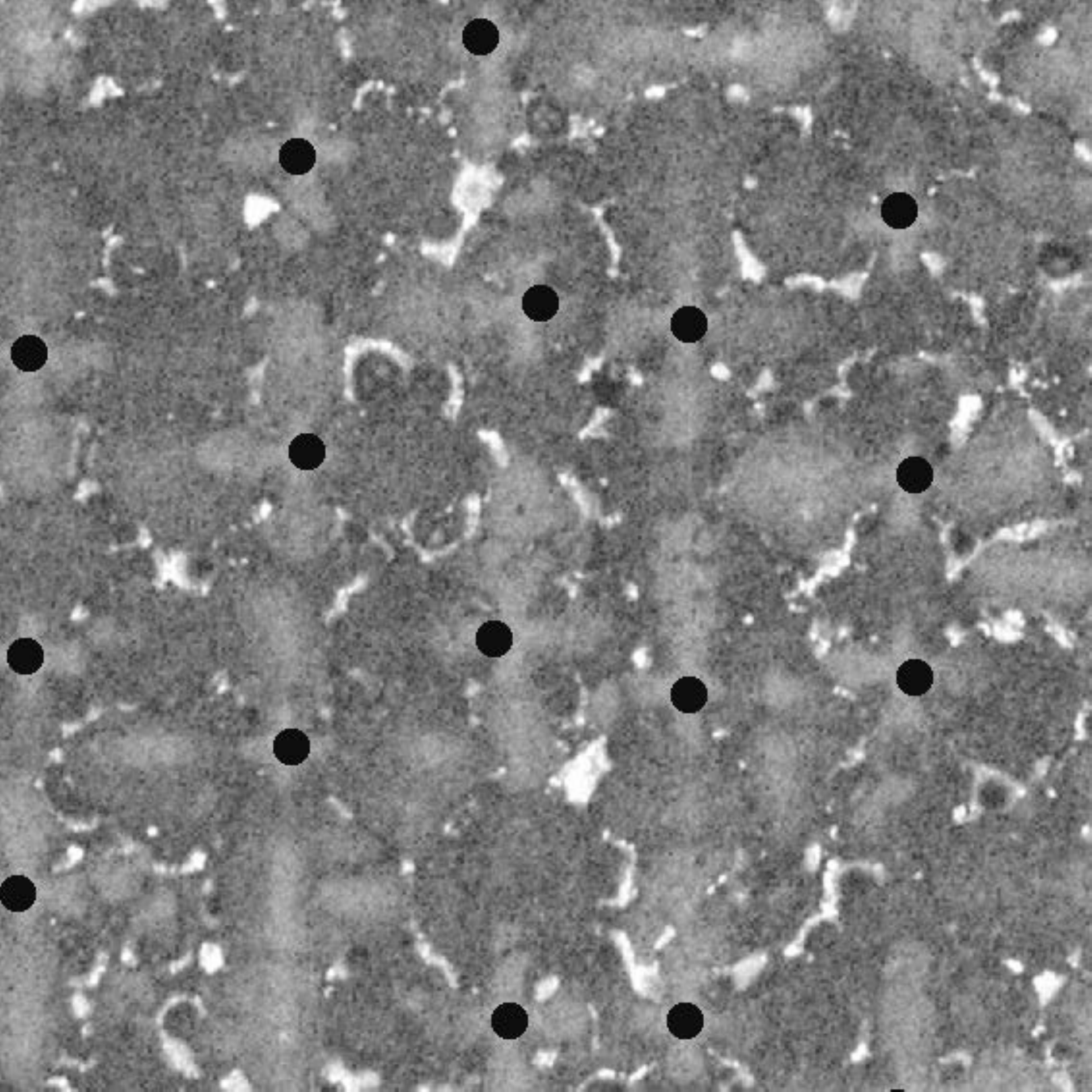}
                \caption{Original image}
	\label{subimages_a}
        \end{subfigure}%
\quad
        \begin{subfigure}[b]{0.31\textwidth}
                \includegraphics[width=\textwidth]{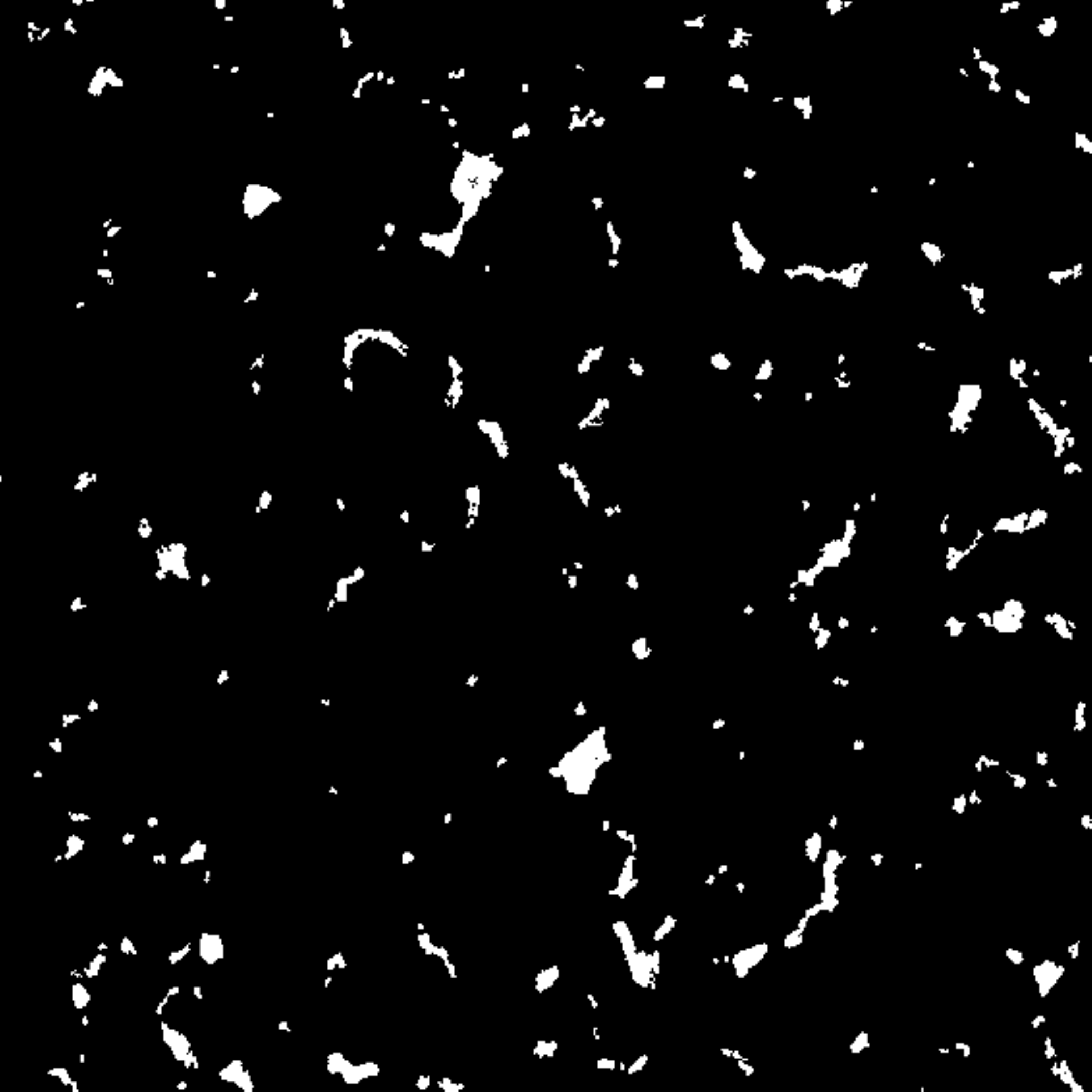}
                \caption{Segmented image}
	\label{subimages_b}
        \end{subfigure}
\quad
        \begin{subfigure}[b]{0.31\textwidth}
                \includegraphics[width=\textwidth]{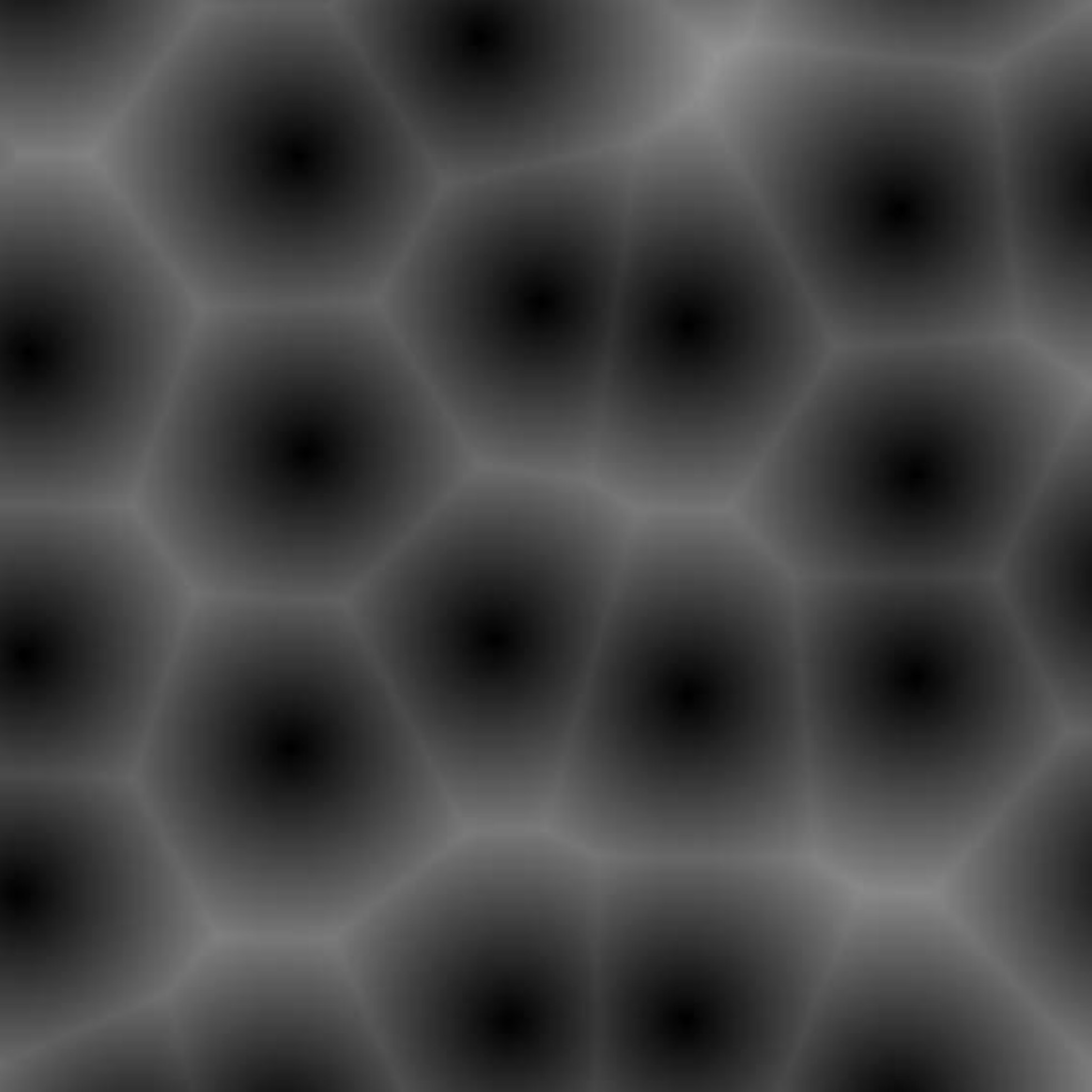}
                \caption{Euclidean map}
	\label{subimages_c}
        \end{subfigure}
\caption[]{(a) A 1 mm x 1 mm subregion from Figure \ref{sx_nickel} is shown along with two corresponding images of the same area: (b) a binary image with segmented eutectic particles (white) and (c) a Euclidean distance map from the dendrite core centroids, where lighter intensity refers to further distances from the dendrite cores.}
\label{subimages}
\end{figure}

The eutectic particles in Figure \ref{sx_nickel} were segmented using the following process.  The particles were segmented by first leveling the intensity of the micrograph using a cubic polynomial with interaction terms.  This step ensured that there wasn't a shift in contrast from one side of the micrograph to the other (due to uneven etching, etc.).  The threshold intensity was then selected by maximizing the difference in the mean intensity between the two distributions (eutectic particle and matrix).  Then, a size threshold was enforced by discarding eutectic particles with less than 5 pixels (1 pixel $\sim{1.7}$ $\mu$m, i.e., 5 pixels = $15.2$ $\mu$m$^2$).  As an example of the segmentation, Figure \ref{subimages_a} shows a 1 mm x 1 mm region from Figure \ref{sx_nickel} and Figure \ref{subimages_b} shows the corresponding binary image of the segmented eutectic particles (in white).  

The Euclidean distance to the nearest dendrite core and the nearest Voronoi vertex was then calculated for each pixel within the micrograph.  The Euclidean distance is the distance from each pixel to the nearest feature, which in this case is either the centroids of the dendrite cores or the Voronoi vertices, and this metric is repeated over all pixels within the image to create a map.  As an example, Figure \ref{subimages_c} shows the Euclidean distance map for the same 1 mm x 1 mm area utilizing the dendrite core centroids identified in Figure \ref{sx_nickel}.  The darker intensity indicates closer Euclidean distances to the dendrite core and the lightest pixels between the dendrite cores actually correspond to the boundaries of the Voronoi tesselation.  

\begin{figure}[bht!]
        \centering
        \begin{subfigure}[b]{0.475\textwidth}
                \centering
	\includegraphics[width=\textwidth]{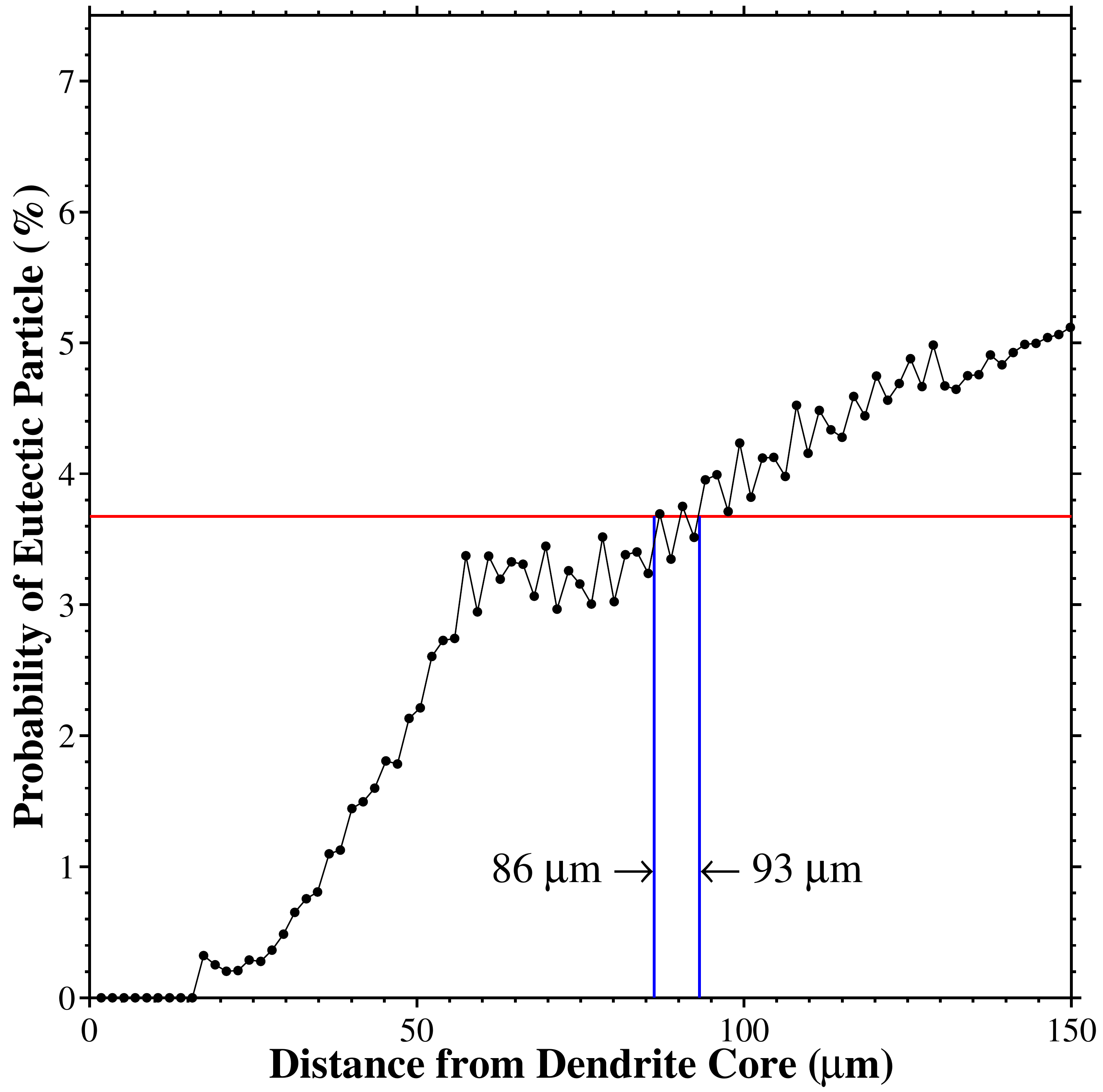}\llap{\raisebox{5cm}{\includegraphics[height=2cm]{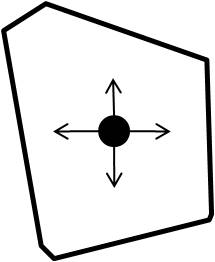}}}
                \caption{}
\label{pdf_a}
        \end{subfigure}%
\quad
        \begin{subfigure}[b]{0.475\textwidth}
                \centering
                \includegraphics[width=\textwidth]{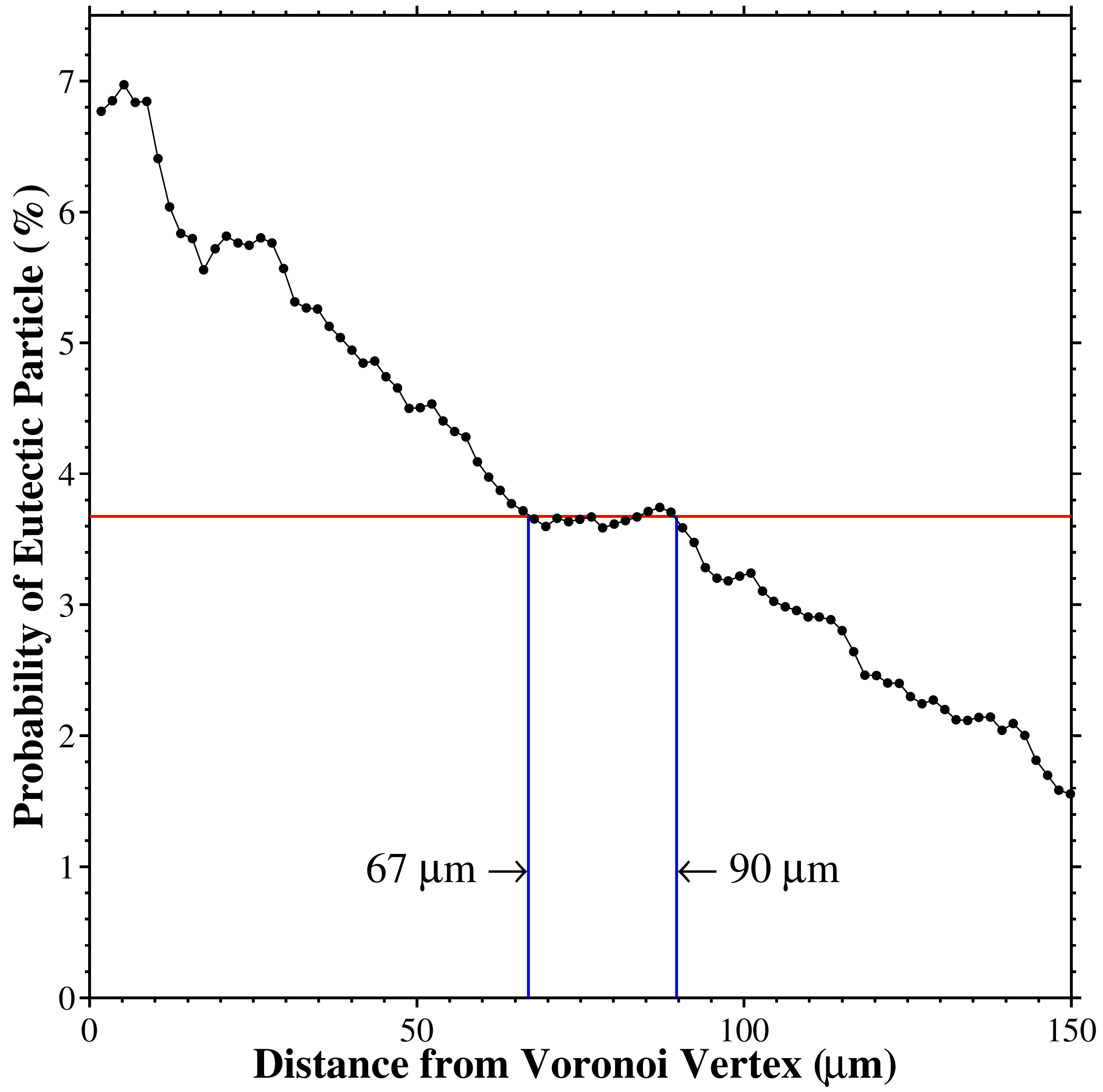}\llap{\raisebox{4.8cm}{\includegraphics[height=2.25cm]{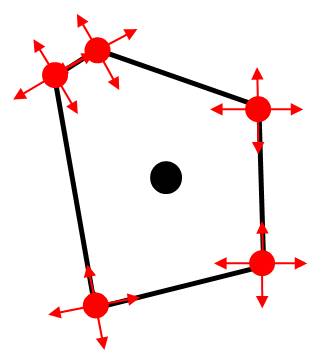}}}
                \caption{}
\label{pdf_b}
        \end{subfigure}
\caption[]{The probability of a eutectic particle as a function of the distance to (a) the nearest dendrite core or (b) the nearest Voronoi vertex.  The inset schematic shows the refence point(s) for the Euclidean distance in each plot.}
\label{pdf}
\end{figure}

The probability of encountering (or forming) a eutectic particle can then be calculated as a function of this Euclidean distance from the nearest dendrite core or the Voronoi vertex, as shown in Figure \ref{pdf}.  Based on the image segmentation, the area fraction of eutectic particles in Figure \ref{sx_nickel} is 3.6\% and is shown as a red line in Figure \ref{pdf}.  The pixels lying within 100 $\mu$m of the image boundaries were excluded to eliminate the possibility that dendrite cores just outside of the field of view could affect the statistics.  As can be seen from Fig.~\ref{pdf_a}, the left (right) blue line indicates the distance whereby all distances below (above) have a probability of having a eutectic particle that is lower (higher) than the global area fraction (red line), i.e., it is less (more) favorable for a eutectic particle to form.  The transition distance of eutectic particle favorability is between 86-93 $\mu$m, i.e., approximately ${1/3}$ of the primary dendrite arm spacing (248.4 $\mu$m).  This plot shows that it is not favorable for eutectic particles to form close to the primary dendrite core.  

Figure \ref{pdf_b} is a similar plot as a function of distance from the vertices of the Voronoi tessellation (see schematic).  This plot was generated in a similar manner to Figure \ref{pdf_a}; a Euclidean distance map was first formed from the Voronoi vertices, then the boundary pixels within 100 $\mu$m of the image boundaries were excluded, etc.  There is an increased occurrence of eutectic particles at vertices, regardless of their distance from the dendrite core.  This observation (along with the fact that the probability of occurence is higher than in Figure \ref{pdf_a} by almost 2\%) suggests that solute is forced near the Voronoi vertices, thereby increasing the probability of eutectic particle occurrence.  The transition distance in this plot is between 67-90 $\mu$m, i.e., at approximately ${1/3}$ of the primary dendrite arm spacing.  While this analysis shows the preferential formation of eutectic particles based on the local distances, correlation with the size of particles is also important.

\begin{figure}[bht!]
        \centering
        \begin{subfigure}[b]{0.475\textwidth}
                \centering
	\includegraphics[width=\textwidth]{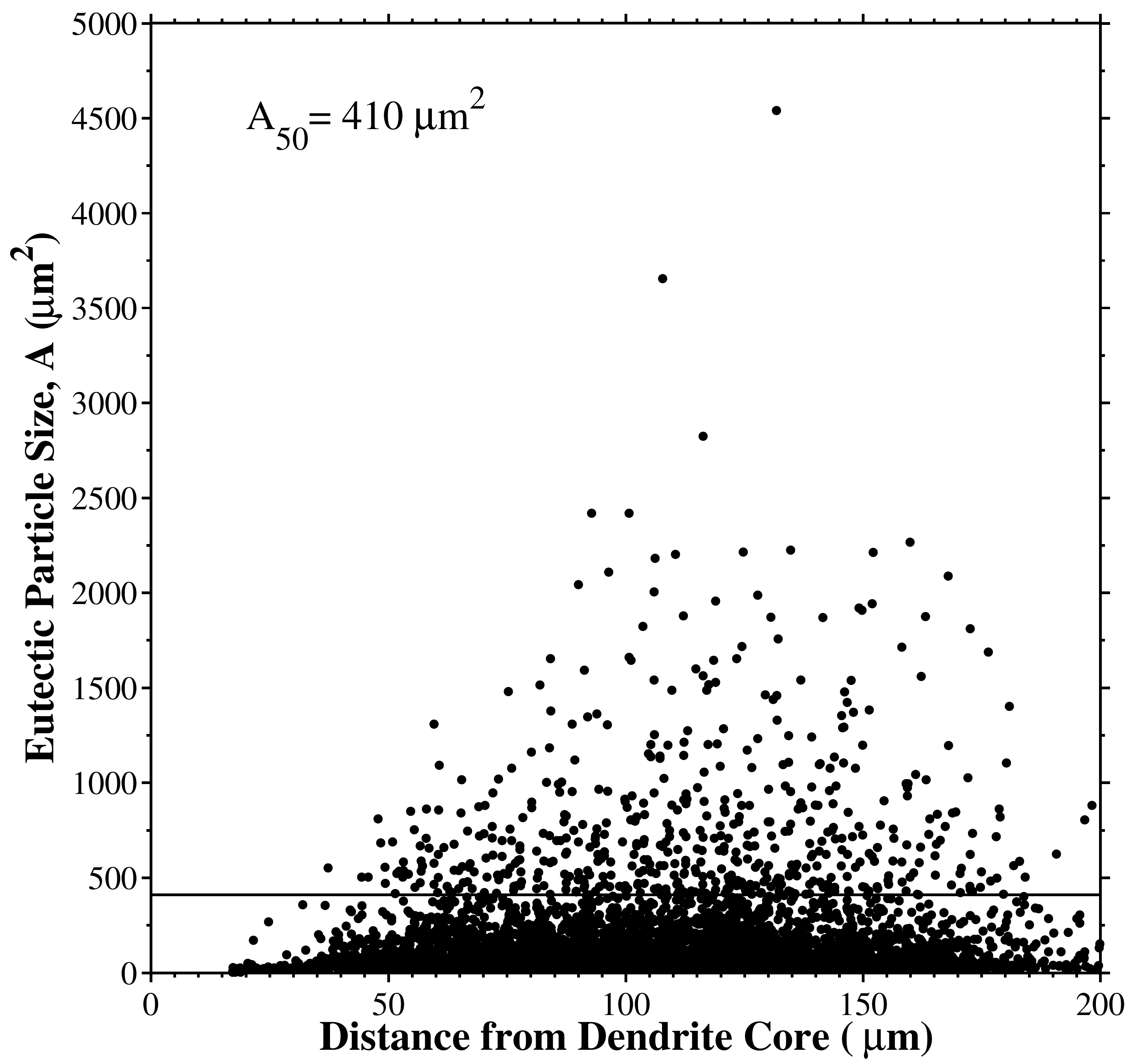}
                \caption{}
\label{a50_a}
        \end{subfigure}%
\quad
        \begin{subfigure}[b]{0.475\textwidth}
                \centering
	\includegraphics[width=\textwidth]{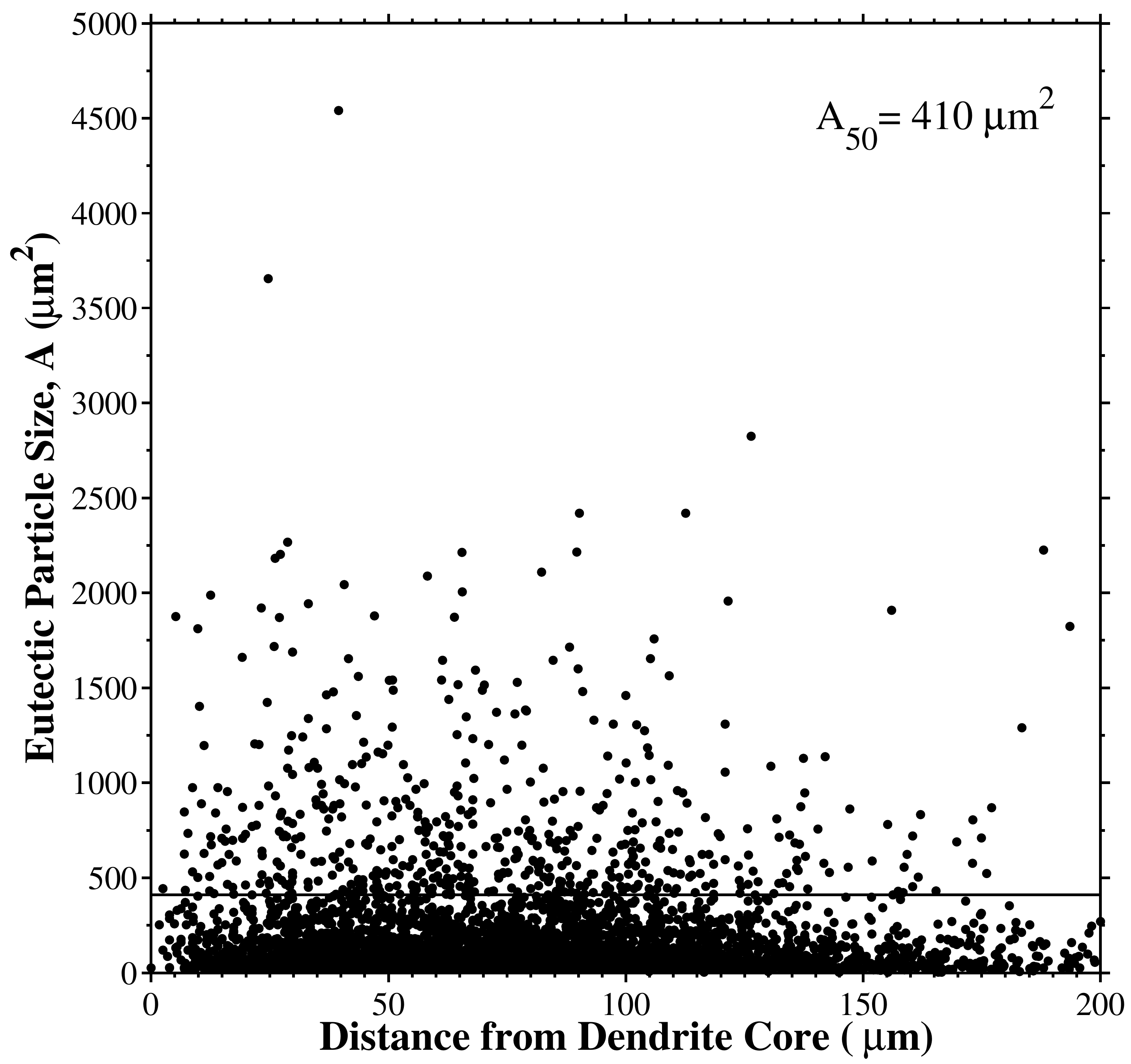}
                \caption{}
	\label{a50_b}
        \end{subfigure}
\caption[]{The eutectic particle size as a function of the distance to (a) the nearest dendrite core or (b) the nearest Voronoi vertex.  The distance for each particle is the distance for the particle centroid.  The 50$^{th}$-percentile area, $A_{50}=410$ $\mu{m}^2$, refers to the particle size where 50\% of the eutectic particle area lies above/below this size.}
\label{a50}
\end{figure}

\begin{figure}[bht!]
        \centering
        \begin{subfigure}[b]{0.475\textwidth}
           \centering
	\includegraphics[width=\textwidth]{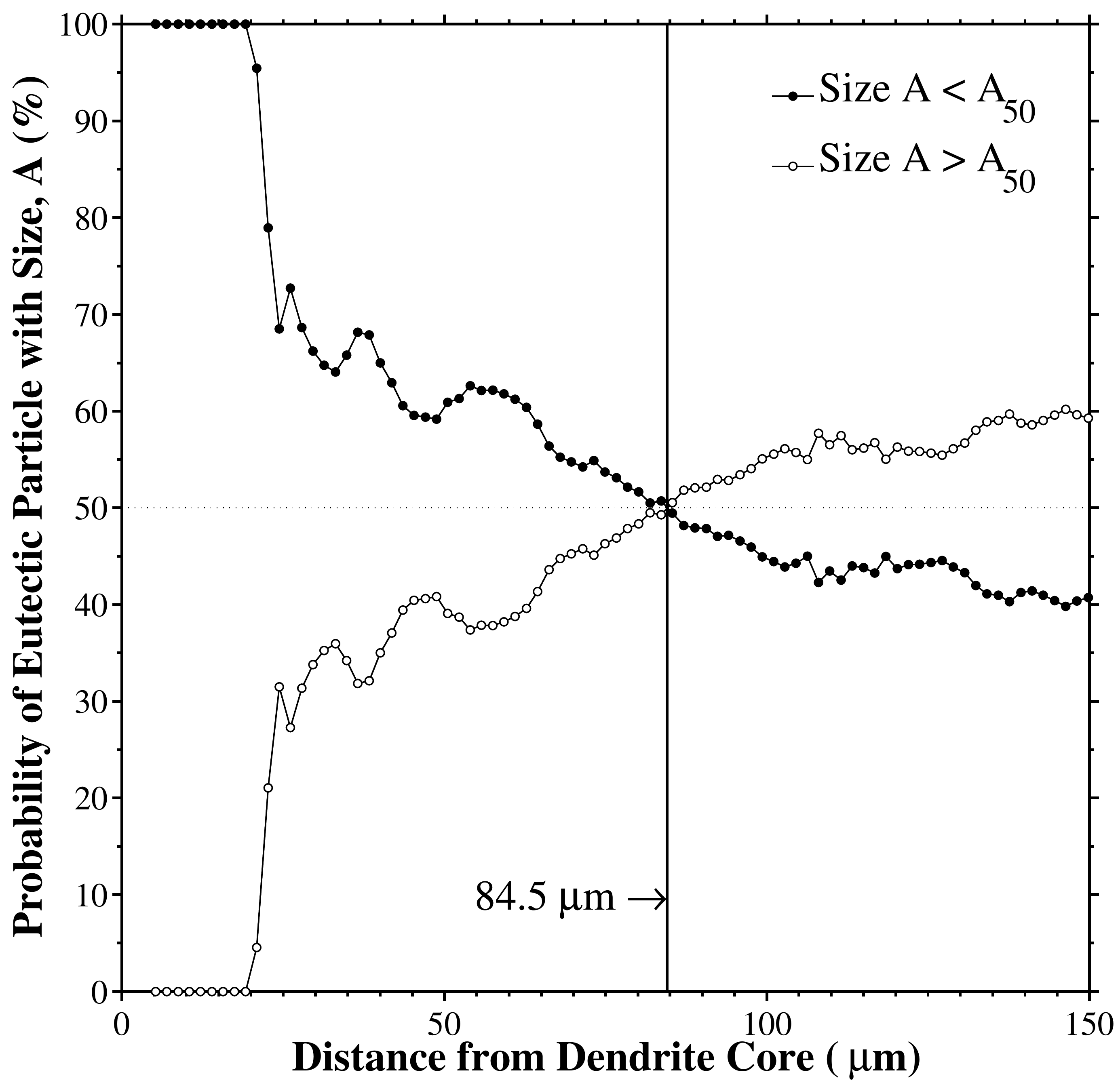}
           \caption{}
	\label{size_a}
        \end{subfigure}%
\quad
        \begin{subfigure}[b]{0.475\textwidth}
                \centering
	\includegraphics[width=\textwidth]{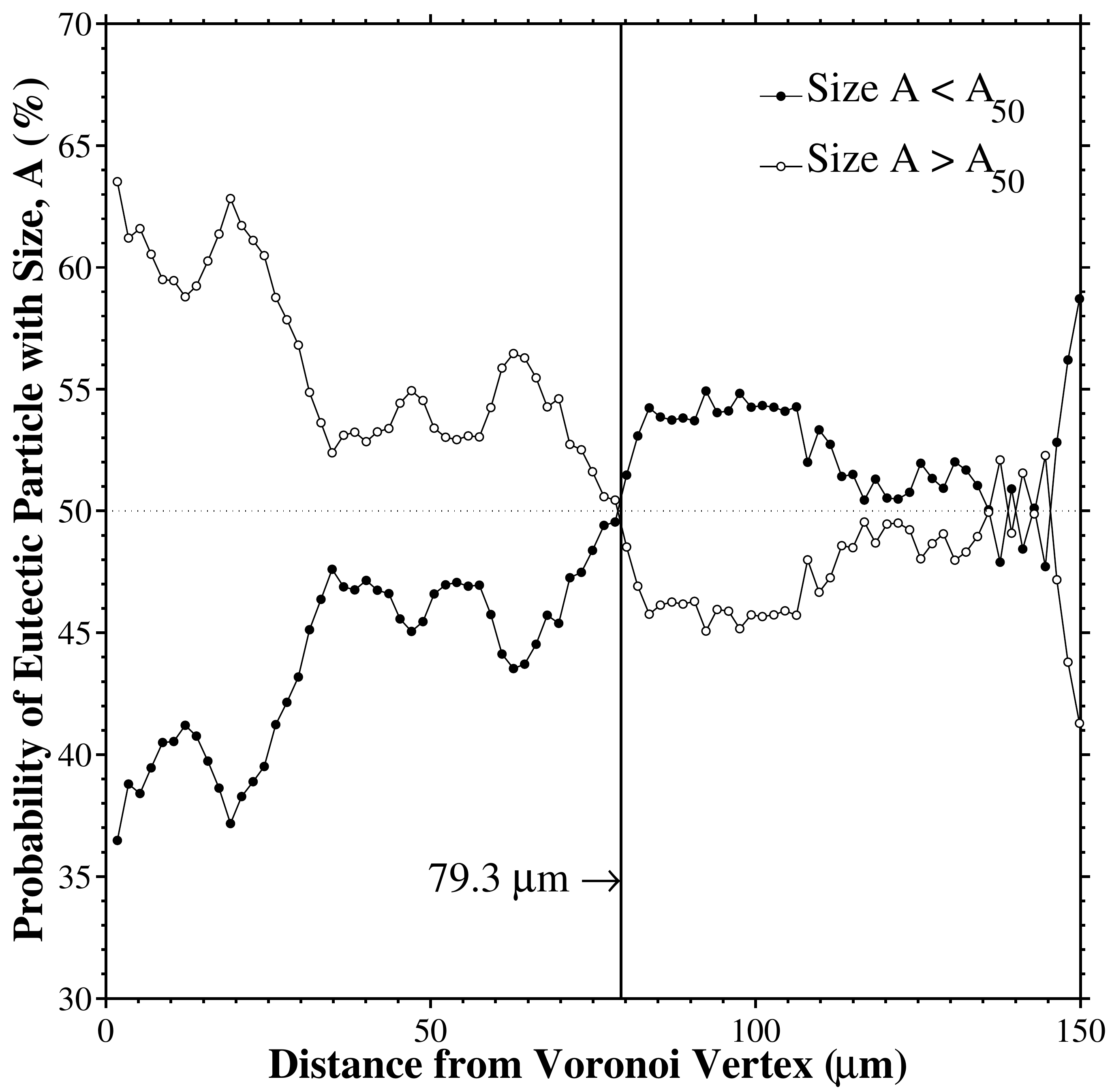}
                \caption{}
	\label{size_b}
        \end{subfigure}

\caption[]{The probability of a eutectic particle of a certain size occurring as a function of the distance to (a) the nearest dendrite core or (b) the nearest Voronoi vertex.  Two particle sizes are considered: particle sizes below and above the 50$^{th}$-percentile area $A_{50}$.  The solid line denotes the distance at which the probability functions first intercept, indicating a transition fromthe favorability of small particles to large particles (in \ref{size_a}) or vice versa (in \ref{size_b}).}
\label{size}
\end{figure}

The eutectic particle size may be correlated with the distance from the dendrite cores or Voronoi vertices as well.  Figure \ref{a50} shows the eutectic particle size as a function of the distance from the nearest dendrite core and the nearest Voronoi vertice.  The solid line shows the 50$^{th}$-percentile area, $A_{50}=410$ $\mu{m}^2$, which refers to the eutectic particle size where 50\% of the eutectic particle area lies above/below this size.  There is a noticeable tendency for the larger particles ($A>A_{50}$) to form further away from the dendrite core and closer to the Voronoi vertices, while the smaller eutectic particles ($A<A_{50}$) can form at all distances.  However, it is difficult to quantitatively tell from the following plot what the preference is for smaller or larger particles as a function of distance.  Therefore, to further quantify this relationship with respect to the size of the particles, the probability associated with a eutectic particle pixel belonging to either a small or large particle is calculated in Figure \ref{size}.  Interestingly, in Figure \ref{size_a}, at distances closer to the dendrite cores, there is a clear preference for smaller particles ($A<A_{50}$) to form over larger particles ($A>A_{50}$).  At a distance of 84.5 $\mu{m}$ ($\sim{1/3}$ PDAS), as denoted by the solid line, there is a crossover in the probability function and larger particles are statistically favored to form over smaller particles.  In the case of distances from the Voronoi vertices, there is a similar behavior except that \textit{larger} particles are favored at smaller distances (closer to Voronoi vertices).  The crossover in the probability functions occurs at 79.3 $\mu{m}$ ($\sim{1/3}$ PDAS again).  At distances greater than this, there is not as definitive of a trend as with the dendrite cores, i.e., in some cases, there is a greater probability for smaller particles to form and, in some cases, for larger particles to form.  This lack of a well-defined trend at larger distances may be caused by the fact that these larger distances could lie close or far away from the dendrite core, further obscuring the trend.  Clearly, the distance from the dendrite cores and, hence, the local primary dendrite arm spacing affect the probabilities of interdendritic particles to form, though.  In a similar manner, it is anticipated that a similar relationship may be associated with shrinkage porosity, gas porosity, and other interdendritic defects.

\section{Conclusions}
In summary, characterizing the primary dendrite arm spacing in directionally-solidified microstructures is an important step for developing process-structure-property relationships by enabling the quantification of (i) the influence of processing on microstructure and (ii) the influence of microstructure on properties.  Thin-walled directionally-solidified structures (e.g., a turbine blade) require new approaches for characterizing the dendrite arm spacing and the microstructure.  In this work, we utilized a new Voronoi-based approach for spatial point pattern analysis that was applied to an experimental dendritic microstructure.  This technique utilizes a Voronoi tessellation of space surrounding the dendrite cores to determine nearest neighbors and the local primary dendrite arm spacing.  In addition, we compared this technique to a recent distance-based technique, the Warnken--Reed method, and a modification to this using Voronoi tesselations, along with the minimal spanning tree method.  Moreover, a convex hull-based technique was used to include edge effects for such techniques, which can be important for thin specimens.  These methods were used to quantify the distribution of local primary dendrite arm spacings as well as their spatial distribution for an experimental directionally-solidified superalloy micrograph.  Last, eutectic particles were segmented to correlate distances from dendrite cores and Voronoi vertices to the occurence and size of these interdendritic features.  Interestingly, with respect to the distance from the dendrite core, it was found that there is a greater probability of occurence of large eutectic particles ($>410$ $\mu$m) over small particles at distances greater than approximately ${1}/{3}$ of the bulk-measured primary dendrite arm spacing.  In conclusion, this systematic study of the different techniques for quantifying local primary dendrite arm spacings, and their effect on microstructure, can be an important step for correlating with both processing and properties in single crystal nickel-based superalloys.

\section*{Acknowledgments}
MAT would like to acknowledge AFOSR for support for this research through contract FA9550-12-1-0135 (PM: Dr. David Stargel, AFOSR/RSA). MAT would like to acknowledge support from the U.S. Army Research Laboratory (ARL) administered by the Oak Ridge Institute for Science and Education through an interagency agreement between the U.S. Department of Energy and ARL.

\bibliographystyle{unsrt}

\end{document}